\documentclass{rQUF2e}
\usepackage{graphicx}
\usepackage{cite}
\usepackage{natbib}
\usepackage{subfigure}
\usepackage{amsmath}

\begin{document}

\title{Emergence of statistically validated financial intraday lead-lag relationships}

\author{Chester Curme$^{1}$, 
Michele Tumminello$^{2}$, Rosario N. Mantegna$^{3,4}$,\\
H. Eugene Stanley$^{1}$, and Dror Y. Kenett$^{1}$}

\affil{$1$ Center for Polymer Studies and Department of Physics,\\
  Boston University, Boston MA, USA\\
$2$ Department of Statistical and Mathematical Sciences ``Silvio Vianelli'' \\ University of Palermo, Palermo, Italy\\
$3$ Center for Network Science and Department of Economics, Central European University, Nador utca 9, 1051 Budapest, Hungary\\
$4$ Department of Physics and Chemistry, University of Palermo, Palermo, Italy
}

\maketitle

\begin{abstract}
According to the leading models in modern finance, the presence of
intraday lead-lag relationships between financial assets is negligible
in efficient markets. With the advance of technology, however, markets
have become more sophisticated. To determine whether this has
resulted in an improved market efficiency, we investigate whether
statistically significant lagged correlation relationships exist in
financial markets. We introduce a numerical method to statistically
validate links in correlation-based networks, and employ our method to study
lagged correlation networks of equity returns in financial markets. Crucially, our statistical validation of lead-lag relationships accounts for multiple hypothesis testing over all stock pairs. In
an analysis of intraday transaction data from the periods 2002--2003 and
2011--2012, we find a striking growth in the networks as we increase the
frequency with which we sample returns. We compute how the number of
validated links and the magnitude of correlations change with increasing
sampling frequency, and compare the results between the two data
sets. Finally, we compare topological properties of the directed correlation-based
networks from the two periods using the in-degree and out-degree
distributions and an analysis of three-node motifs. Our analysis
suggests a growth in both the efficiency and instability of financial
markets over the past decade.
\end{abstract}

\section{Introduction}
Modern financial markets have developed lives of their own. This fact
makes it necessary that we not only monitor financial markets as an
``auxiliary system'' of the economy, but that we develop a methodology for
evaluating them, their feedback on the real economy, and their effect on
society as a whole \citep{hall2010does,cecchetti2012reassessing}. The
events of the recent past have clearly demonstrated that the everyday
life of the majority of the world's population is tied to the well-being
of the financial system. Individuals are invested in stock markets
either directly or indirectly, and shocks to the system (be they
endogenous or exogenous) have an immense and immediate impact. Thus the
need for a robust and efficient financial system is becoming stronger
and stronger. These two critical concepts have been discussed and
heatedly debated for the past century, with the efficient market
hypothesis (EMH) in the center of the debate.

The EMH
\citep{tobin1969general,malkiel1970efficient} stipulates that all
available information (or only past prices in the weak variant of the
hypothesis) is already reflected in the current price and it is
therefore not possible to predict future values in any statistical
method based on past records \citep{malkiel2003efficient}. The EMH
has been questioned by applying statistical tests to NYSE returns
\citep{lo1988stock,shmilovici2003using} in which the authors formulated
the problem equivalent to the EMH, and showed by contrast that an
efficient compression algorithm they proposed was able to utilize
structure in the data---which would not be possible if the hypothesis
were in fact true. The possibility for such compression suggests the
data must be somehow structured. This encourages us to explore methods
of modeling and exploring this structure in ways that can be applied to
real-world markets.

Many efforts have thus been devoted to uncovering the true nature of the
underlying structure of financial markets. Much attention has been given
to understanding correlations in financial markets and their dynamics,
for both daily
\citep{mantegna1999hierarchical,gopikrishnan2000scaling,cizeau2001correlation,forbes2002no,campbell2008increasing,podobnik2008detrended,carbone2009detrending,aste2010correlation,pollet2010average,Kenett2012e,Kenett2012a}
and intra-day time scales
\citep{bonanno2001high,borghesi2007emergence,tumminello2007correlation,munnix2010impact}. More
recently, other measures of similarity have been introduced, such as
Granger-causality analysis \citep{billio2012econometric} and partial
correlation analysis \citep{kenett2010b}, both of which aim to quantify
how the behavior of one financial asset provides information about the
behavior of a second asset. For these different measures of co-movement
in financial markets, however, the main question that remains is how to
uncover underlying meaningful information.

An analysis of synchronous correlations of equity returns has shown that
a financial market usually displays a nested structure in which all the
stock returns are driven by a common factor, e.g., a market index, and
are then organized in groups of like economic activity---such as
technology, services, utilities, or energy---that exhibit higher values
of average pair correlation. Within each group, stocks belonging to the
same sub-sector of economic activity, e.g., ``insurance'' and ``regional
banks'' within the financial sector, show an even higher correlation
degree. Such a structure has been recognized using very different
methods of analysis, ranging from random matrix theory
\citep{laloux2000random,gopikrishnan2001quantifying}, to hierarchical
clustering \citep{mantegna1999hierarchical}, to correlation based
networks \citep{mantegna1999hierarchical,bonanno2003topology,onnela2003dynamic}. The several
methods devised to construct correlation based networks can be grouped
into two main categories: threshold methods and topological/hierarchical
methods. Both approaches start from a sample correlation matrix or, more
generally, a sample similarity measure. Using the threshold method we set
a correlation threshold and construct a network in which any two nodes
are linked if their correlation is larger than the threshold. As we
lower the threshold value we see the formation of groups of stocks
(economic sub-sectors) that progressively merge to form larger groups
(economic sectors) and finally merge into a single group (the
market). The advantage of this approach is that, due to the finite
length of data series, threshold networks are very robust to correlation
uncertainty. The disadvantage of threshold based networks is that it is
difficult to find a single threshold value to display, in a single
network, the nested structure of the correlation matrix of stock returns
(see \citep{kenett2010b}). Topological methods to construct correlation
based networks, such as the minimal spanning tree (MST)
\citep{mantegna1999hierarchical,bonanno2001high,bonanno2003topology,onnela2003dynamic} or the planar
maximally-filtered graph (PMFG) \citep{Tumminello2005}, are based solely
on the ranking of empirical correlations. The advantage of this approach
is that these methods are intrinsically hierarchical and are able to
display the nested structure of stock-return correlations in a financial
market. The disadvantage of this approach is that these methods are less
stable than threshold methods with respect to the statistical
uncertainty of data series, and it is difficult to include information
about the statistical significance of correlations and their ranking
\citep{tumminello17and}. Thus it is a challenge of modern network
science to uncover the significant relationships (links) between the
components (nodes) of the investigated system \citep{
  havlin2012challenges}.

Although much attention has been devoted to the study of synchronous
correlation networks of equity returns (see
\citep{tumminello2010correlation} for a review of the topic),
comparatively few results have been obtained for networks of lagged
correlations \citep{huth2011high}. Neither method of constructing correlation based networks is readily
extendable to the study of directed lagged correlations in a financial
market. The lagged correlations in stock returns are small, even at time
horizons as short as five minutes, and are thus strongly influenced by
the statistical uncertainty of the estimation process. The use of
topological methods to construct a lagged-correlation based network of
stock returns is difficult because they only take into consideration the
ranking of correlations and not their actual values. The result could be
a network in which many links are simply caused by statistical
fluctuations. On the other hand, standard threshold methods are also
difficult to apply because it is difficult to find an appropriate
threshold level and, more importantly, the threshold selected in these
methods is usually the same for all stock pairs. This is a problem if we
want to study lagged correlations because the statistical
significance of a lagged-correlation may depend on the return
distribution of the corresponding pair of stocks, and such distributions might
vary across stocks---a consequence, for example, of the different
liquidity of stocks.

Here we introduce a method for filtering a lagged correlation matrix
into a network of statistically-validated directed links that takes into
account the heterogeneity of stock return distributions. This is done by
associating a $p$-value with each observed lagged-correlation and then
setting a threshold on $p$-values, i.e., setting a level of statistical
significance corrected for multiple hypothesis testing. We apply our
method to describe the structure of lagged relationships between
intraday equity returns sampled at high frequencies in financial
markets. In particular, we investigate how the structure of the network
changes with increasing return sampling frequency, and compare the
results using data from both the periods 2002--2003 and 2011--2012. It should be noted that the two investigated time periods are quite different if we consider that the fraction of volume exchanged by algorithmic trading in the US equity markets has increased from approximately 20\% in 2003 to more than 50\% in 2011. In
both periods we find a large growth in the connectedness of the networks
as we increase the sampling frequency.

The paper is organized as follows. Section 2 introduces the method used
to filter and validate statistically significant lagged correlations
from transaction data. Section~3 analyzes the structure of the resulting
networks and investigates how this structure evolves with changing
return sampling frequency. In Sec.~4 we discuss the application of our
method to the construction of synchronous correlation networks. Finally,
in Sec.~5 we discuss the implications of our results for the efficiency
and stability of financial markets.

\section{Statistically validated lagged correlation networks (SVLCN)}
\label{sec:methods}

\begin{figure}[h]                                      
\begin{center}
\centerline{
 \includegraphics[scale=.6]{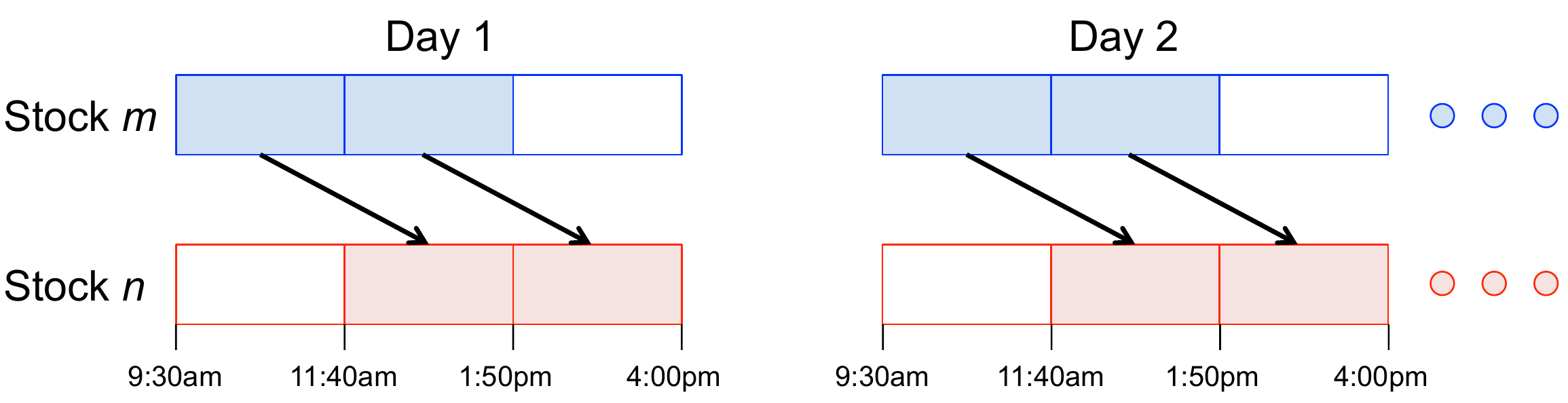}}
 \end{center}
\caption{Schematic of lagged correlation calculation for a time horizon
  $h = 130$ minutes. The sum $C_{m,n}$ is generated using products of
  returns from stocks $m$ and $n$ that are linked by an arrow. We
  consider only time horizons $h$ that divide evenly into the 390 minute
  trading day.}
  \label{fig:lag_schem}
\end{figure}

We begin the analysis by calculating the matrix of logarithmic returns
over given intraday time-horizons. We denote by $p_n(t)$ the most recent
transaction price for stock $n$ occurring on or before time $t$ during
the trading day. We define the opening price of the stock to be the
price of its first transaction of the trading day. Let $h$ be the time
horizon. Then for each stock we sample the logarithmic returns,
\begin{equation}
 r_{n,t} \equiv \log(p_n(t)) - \log(p_n(t-h)),
\end{equation}
every $h$ minutes throughout the trading day, and assemble these time
series as columns in a matrix $R$. We then filter $R$ into two matrices,
$A$ and $B$, in which we exclude returns during the last period $h$ of each
trading day from $A$ and returns during the first period $h$ of each
trading day from $B$. From these data we construct an empirical lagged
correlation matrix $C$ using the Pearson correlation coefficient of
columns of $A$ and $B$,
\begin{equation}
C_{m,n} = \dfrac{1}{T-1} \sum_{i=1}^T\dfrac{(A_{m,i}-\langle A_m\rangle) (B_{n,i}-\langle B_n\rangle)}{\sigma_m\sigma_n},
\label{eqn:corr_matrix}
\end{equation}
where $\langle A_m\rangle$ and $\sigma_m$ are the mean and sample
standard deviation, respectively, of column $m$ of $A$, and $T$ is the
number of rows in $A$ (and $B$). Here we set the lag to be one time
horizon $h$. A schematic of this sum is diagrammed in
Fig.~\ref{fig:lag_schem}. 

The matrix $C$ can be considered a weighted adjacency matrix for a fully
connected, directed graph. To filter the links in this graph according
to a threshold of statistical significance, we apply a shuffling
technique \citep{efron1993introduction}. The rows of
$A$ are shuffled repeatedly without replacement in order to create a
large number of surrogated time series of returns. After each shuffling
we re-calculate the lagged correlation matrix (\ref{eqn:corr_matrix})
and compare this shuffled lagged correlation matrix $\widetilde{C}$
to the empirical matrix $C$. For each shuffling we thus have an
independent realization of $\widetilde{C}$. We then construct the
matrices $U$ and $D$, where $U_{m,n}$ is the number of realizations for
which $\widetilde{C}_{m,n} \geq C_{m,n}$, and $D_{m,n}$ is the number of
realizations for which $\widetilde{C}_{m,n} \leq C_{m,n}$.

From matrix $U$ we associate a one-tailed $p$-value with all positive
correlations as the probability of observing a correlation that is equal
to or higher than the empirically-measured correlation. Similarly, from
$D$ we associate a one-tailed $p$-value with all negative
correlations. In this analysis we set the threshold at $p=0.01$. We must
adjust our statistical threshold, however, to account for multiple
comparisons. We use the conservative Bonferroni correction for a given
sample size of $N$ stocks. For example, for $N=100$ stocks the corrected
threshold will be $0.01/N^2 = 10^{-6}$. We thus construct $10^6$
independently shuffled surrogate time series. If $U_{m,n}=0$ we can
associate a statistically-validated positive link from stock $m$ to
stock $n$ ($p=0.01$, Bonferroni correction). Likewise, if $D_{m,n}=0$ we
can associate a statistically-validated negative link from stock $m$ to
stock $n$. In this way we construct the Bonferroni network
\citep{tumminello2011statistically}. In Appendix A we discuss the
probability that using our approximated method we will wrongly indentify
a link as statistically significant (i.e., have a false positive).

For the sake of comparison, for each time horizon $h$ we also construct
the network using $p$-values corrected according to the false discovery
rate (FDR) protocol \citep{benjamini1995controlling}. This correction is
less conservative than the Bonferroni correction and is constructed as
follows. The $p$-values from each individual test are arranged in
increasing order ($p_1 < p_2 < \dots < p_{N^2}$), and the threshold is
defined as the largest $k$ such that $p_k < k~0.01/N^2$. In the FDR
network our threshold for the matrices $U$ or $D$ is thus not zero but
the largest integer $k$ such that $U$ or $D$ has exactly $k$ entries
fewer than or equal to $k$. From this threshold we can filter the links
in $C$ to construct the FDR network
\citep{tumminello2011statistically}. We note that the Bonferroni network is a
subgraph of the FDR network.

Because we make no assumptions about the return distributions, this
randomization approach is especially useful in
high-dimensional systems in which it can be difficult to infer the joint
probability distribution from the data \citep{tumminello17and}. We also
impose no topological constraints on the Bonferroni or FDR
networks. This method serves to identify the significant positive and
negative lagged correlation coefficients in a way that accounts for
heterogeneities in relationships between the returns of stocks. An
alternative, but closely related approach would be to construct a
theoretical distribution for correlation coefficients under the null
hypothesis of uncorrelated returns sampled from a given joint
distribution \citep{biroli2007student}. For a desired confidence level,
one could then construct a threshold correlation, beyond which empirical
correlations are validated. Such an approach typically assumes equal
marginal distributions for returns, and must fix a uniform correlation
threshold for all relationships. At the expense of computational time,
our method is flexible in that it permits heterogeneities in marginal
distributions. We compare the results of the two approaches in Appendix
B.

\section{Lagged correlation networks in NYSE}

We study and compare two different datasets. The first
dataset comprises returns of 100 companies with the largest market
capitalization on the New York Stock Exchange (NYSE) during the period
2002--2003 (501 trading days), which was investigated in
\citep{tumminello2007correlation}. For the second dataset we consider
returns during the period 2011--2012 (502 trading days) of 100 companies
with the largest market capitalization on the NYSE as of December 31,
2012 (retrieved from the Trades and Quotes database, Wharton Research
Data Services, http://wrds-web.wharton.upenn.edu/wrds/). Market
capitalization figures were obtained from Yahoo Finance web service
(http://finance.yahoo.com). For each company we obtain intraday
transaction records. These records provide
transaction price data at a time resolution of one second. The stocks
under consideration are quite liquid, helping to control for the problem
of asynchronous transactions and artificial lead-lag relationships due
to different transaction frequencies \citep{de1996price}. We sample
returns at time horizons of 5, 15, 30, 65, and 130 minutes.

We report summary statistics in Table \ref{summary_table}, including the lengths of time series $T$ from equation (\ref{eqn:corr_matrix}), as well as the mean $\langle\rho\rangle$ and
standard deviation $\sigma_\rho$ of synchronous Pearson correlation coefficients between distinct columns of the returns matrix $R$ for
each time horizon $h$. We also show the mean $\langle C_{m,n}\rangle$ and standard deviation $\sigma_C$ of entries in the lagged correlation matrix $C$.

\begin{table}
\centering
\tbl{Summary statistics of 2002-2003 and 2011-2012 datasets.}
{\begin{tabular}{@{}ccccccc}
\toprule
Period&$T$ & $h$ & $\langle \rho \rangle$ & $\sigma_\rho$ & $\langle C_{m,n}\rangle$ & $\sigma_C$ \\
\hline
&38,577 & 5 min. & 0.267 & 0.077 & 0.008 & 0.024 \\
&12,525 & 15 min. & 0.290 &  0.092 & 0.007 & 0.025\\
2002-2003&6,012 & 30 min. & 0.307  & 0.102 & 0.005 & 0.025\\
&2,505 & 65 min. & 0.317  & 0.110 & 0.015 & 0.029 \\
&1002 & 130 min. & 0.327 & 0.115 & 0.022 & 0.038\\
\hline
&38,654 & 5 min. & 0.356 & 0.143 & 0.004 & 0.024 \\
&12,550 & 15 min. & 0.410 &  0.115 & 0.006 & 0.022\\
2011-2012&6,024 & 30 min. & 0.422  & 0.115 & 0.017 & 0.024\\
&2,510 & 65 min. & 0.438  & 0.132 & -0.004 & 0.028 \\
&1004 & 130 min. & 0.451 & 0.126 & -0.019 & 0.034\\
\botrule

\end{tabular}}
\label{summary_table}
\end{table}

\begin{figure}[h]                                      
\begin{center}
\centerline{
 \includegraphics[scale=.4]{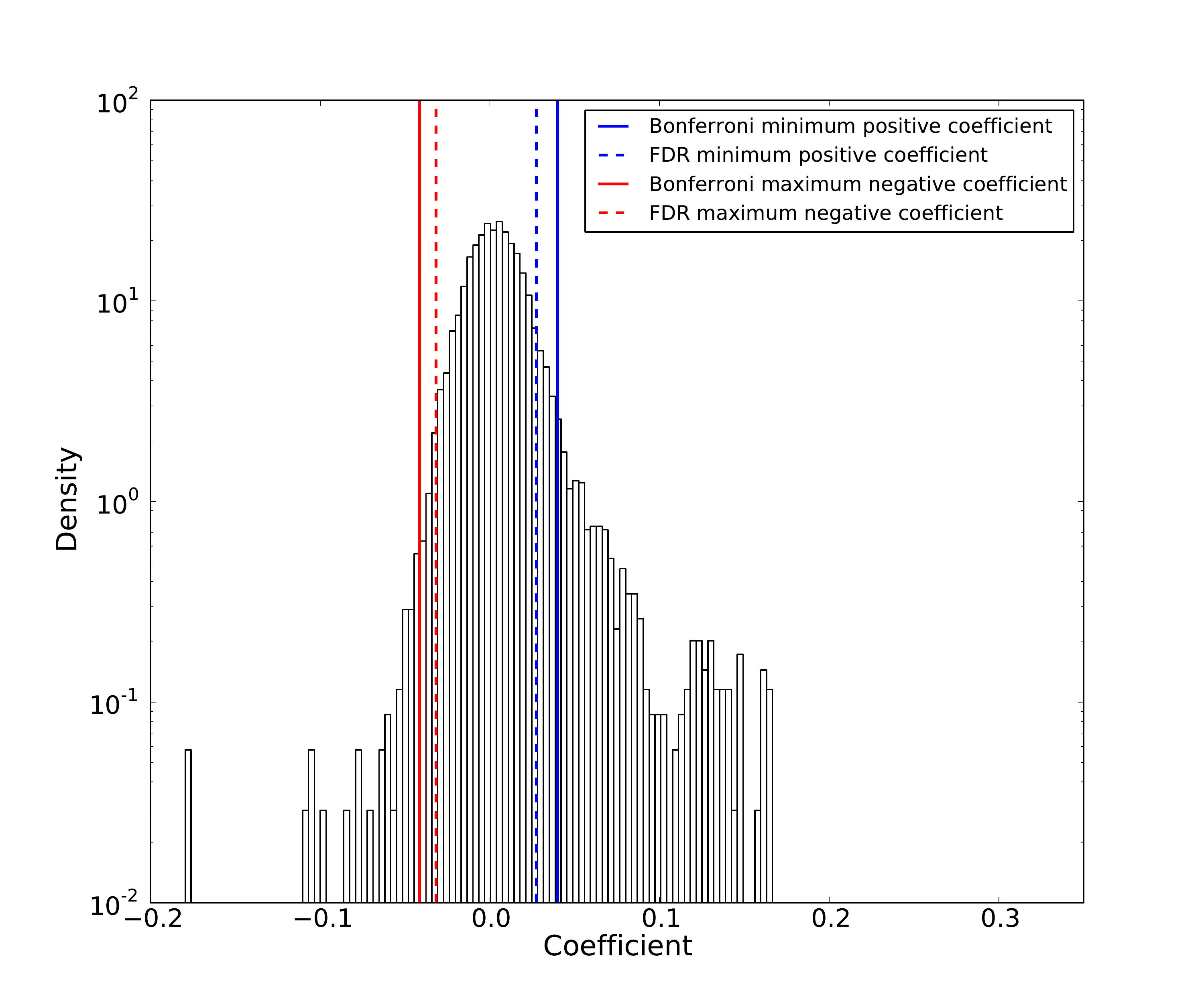}}
 \end{center}
\caption{Distribution of lagged correlation coefficients for all $N=100$
  stocks at a time horizon $h = 15$ minutes. The minimum positive
  coefficients and maximum negative coefficients selected using both
  Bonferroni and FDR filtering procedures are shown. We note that these
  methods select coefficients from the tails of the distribution,
  without fixing a uniform threshold for all pairs of stocks.}
  \label{fig:bounds}
\end{figure}

Figure \ref{fig:bounds} displays bounds on the positive and negative
coefficients selected by this method for both Bonferroni and FDR
networks at a time horizon of $h = 15$ minutes.

\begin{figure}[h]
    \begin{center}
    \centerline{
        \subfigure[$h=130$ minutes]{%
            \label{fig:h130}
            \includegraphics[width=0.33\textwidth]{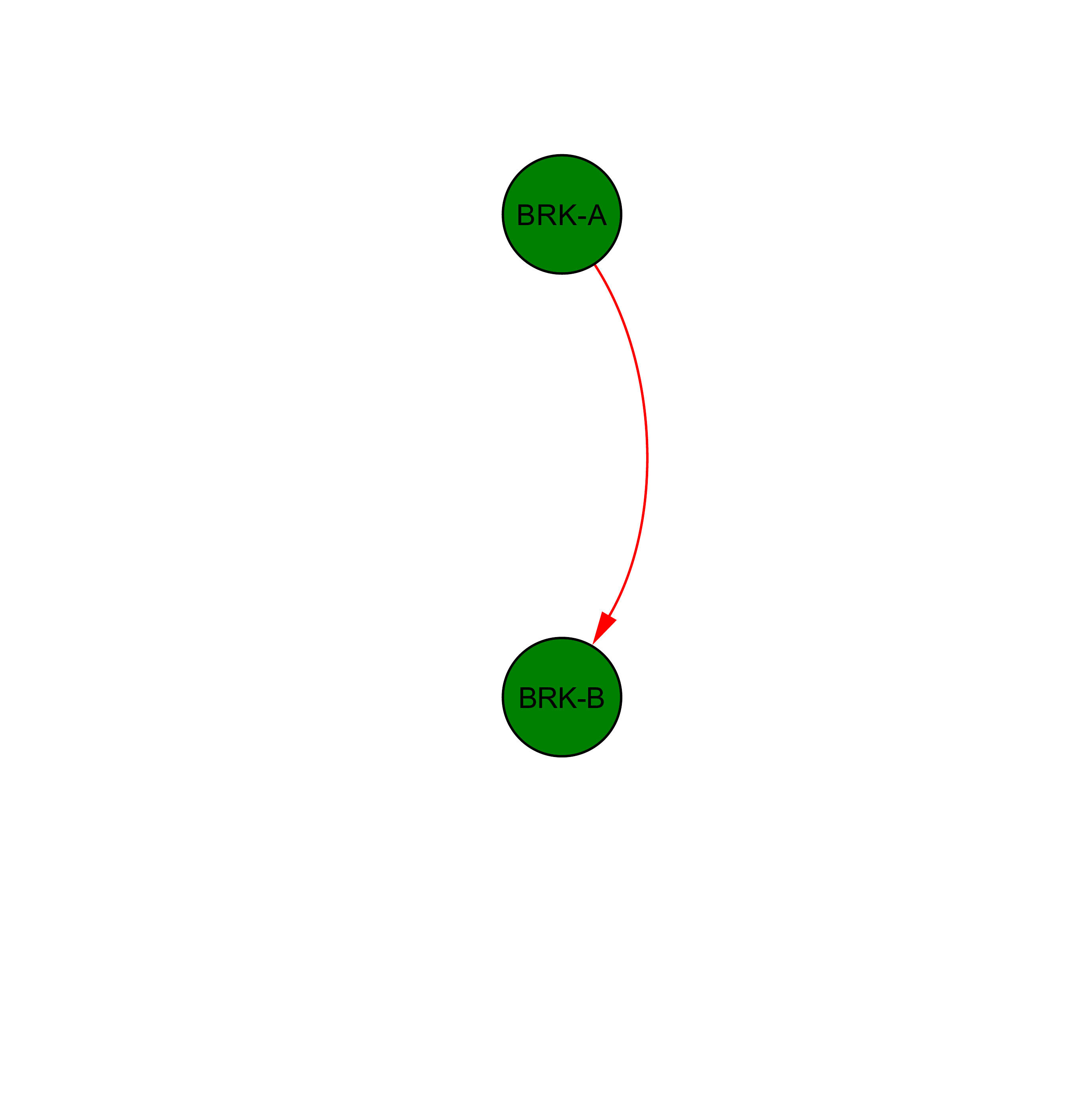}
        }
        \subfigure[$h=65$ minutes]{%
            \label{fig:h65}
            \includegraphics[width=0.33\textwidth]{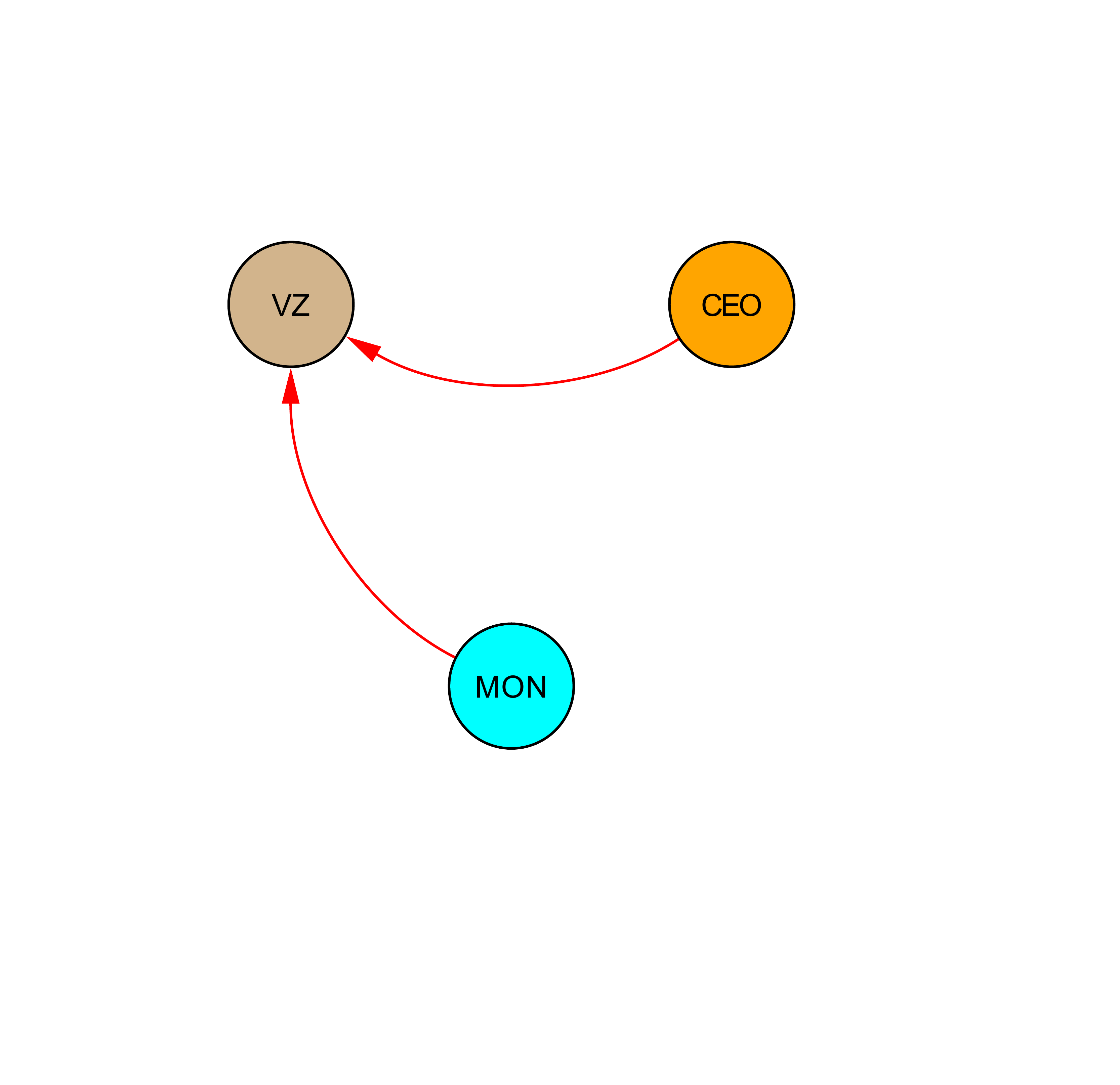}
            }
            \subfigure[$h=30$ minutes]{%
            \label{fig:h30}
            \includegraphics[width=0.33\textwidth]{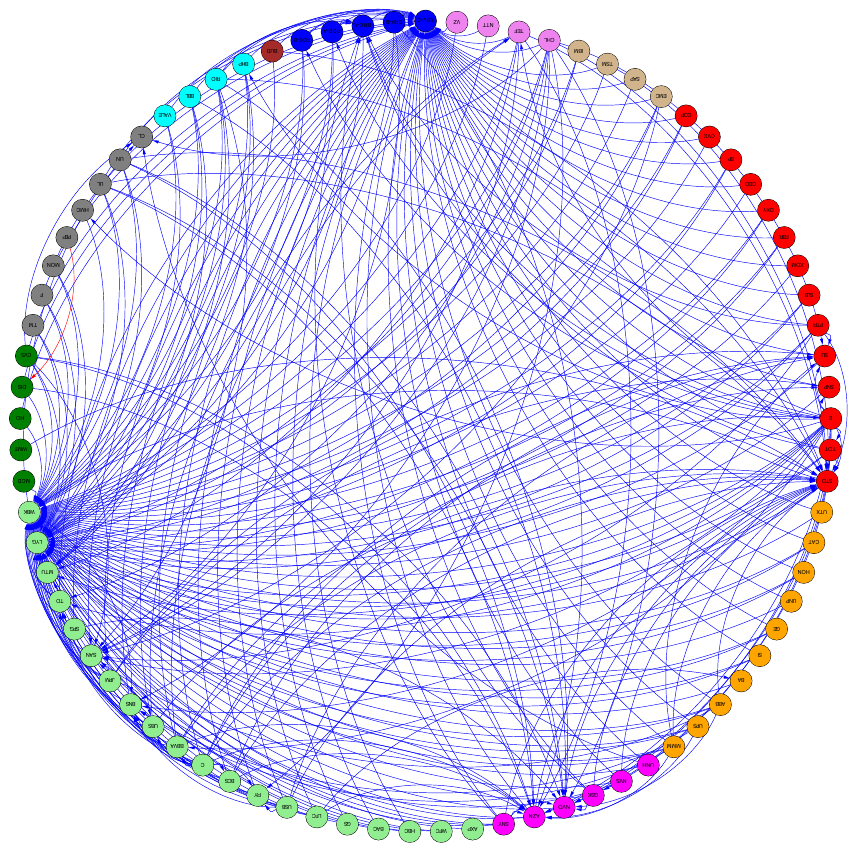}
            }}
            \centerline{
        \subfigure[$h=15$ minutes]{%
            \label{fig:h15}
            \includegraphics[width=0.4\textwidth]{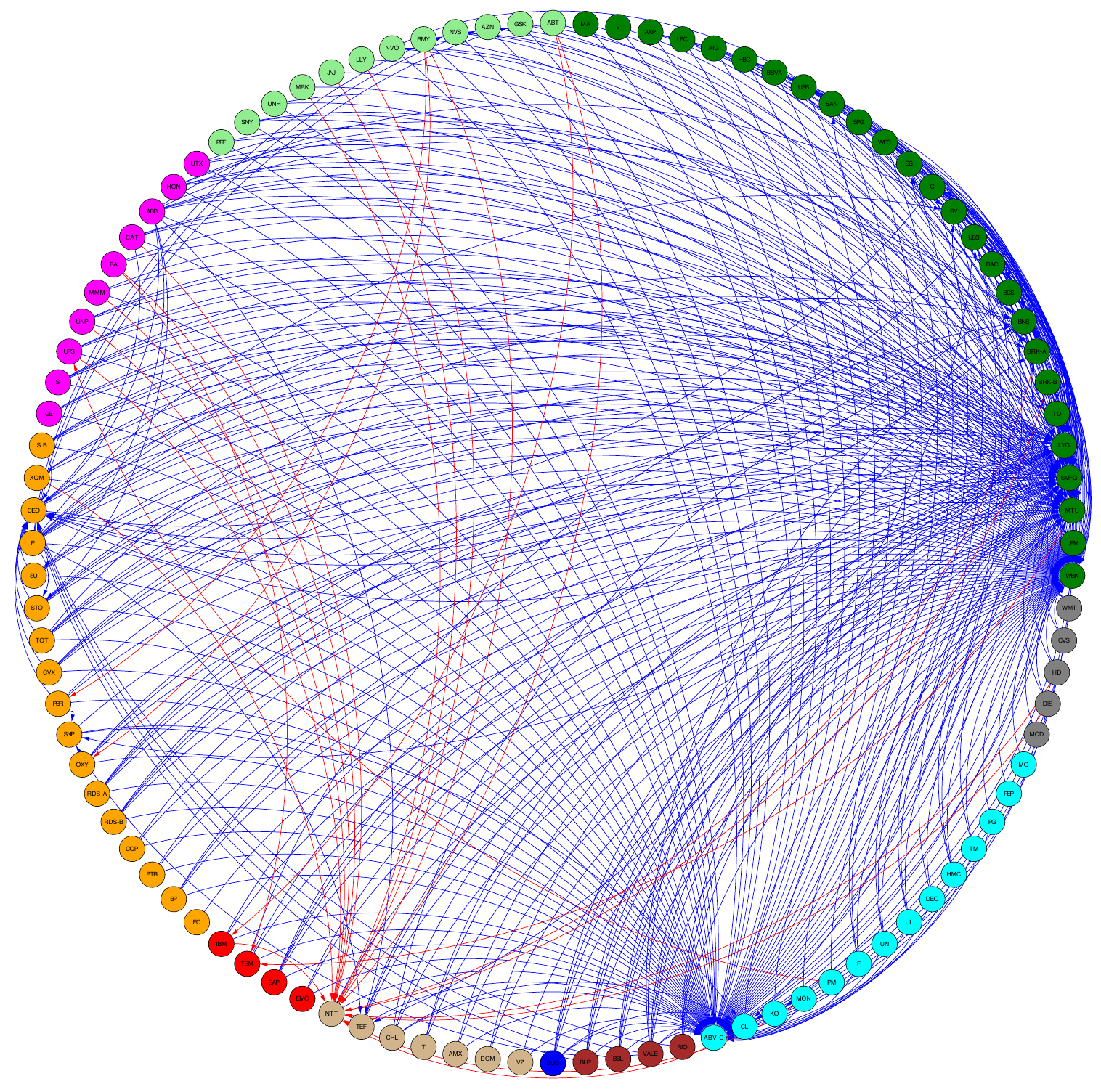}
            }
        \subfigure[$h=5$ minutes]{%
            \label{fig:h5}
            \includegraphics[width=0.4\textwidth]{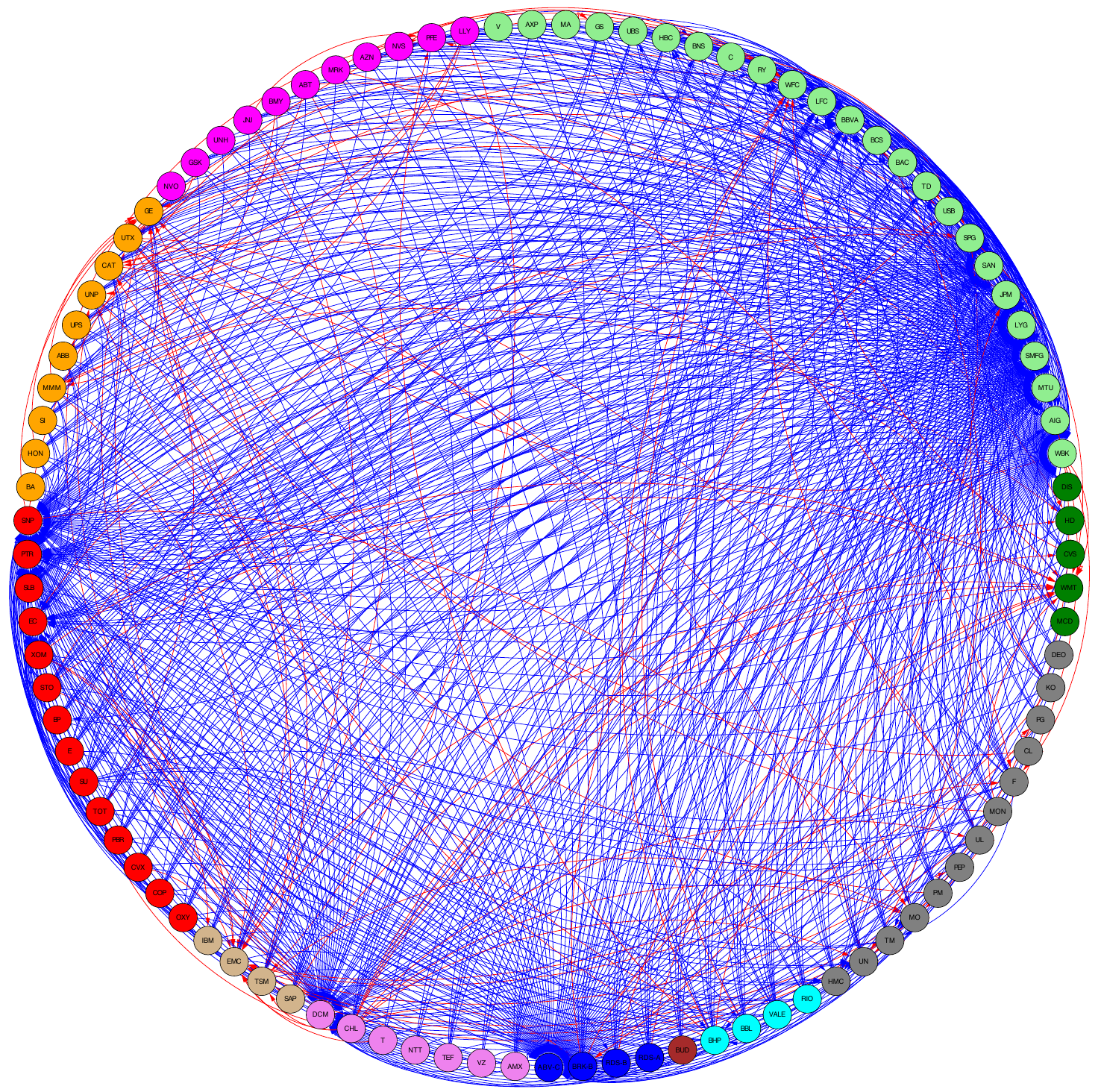}
            }}
    \end{center}
    \caption{ Illustrations of Bonferroni networks constructed from
      statistically-validated lagged correlations for various time
      horizons $h$. Data were obtained from returns of large
      market-capitalization companies on the NYSE in 2011-2012. Nodes
      are colored by industry. Blue links represent positive
      correlations; red links represent negative correlations.  }
    \label{fig:networks}
\end{figure}

In Fig.~\ref{fig:networks} we display plots of each statistically
validated lagged correlation network obtained from the 2011--2012 data
(Bonferroni correction). At time horizons of $h=130$ minutes and $h=65$
minutes we validate one and two links, respectively. It is somewhat
remarkable that we uncover any persistent relationships at such long
time horizons.

We see a striking increase in the number of validated links at small
intraday time horizons, below $h = 30$ minutes in particular. This is
likely due to a confluence of two effects: (i) with decreasing $h$ we
increase the length $T$ of our time series, gaining statistical power
and therefore the ability to reject the null hypothesis; (ii) at small
$h$ we approach the timescales over which information and returns spill
over across different equities. In Appendix C we provide evidence that diminishing the time horizon $h$ reveals more information about the system than is obtained by increasing the time series length $T$ alone.

\begin{figure}[h]
    \begin{center}
	 \centerline{
        \subfigure[Links of positive correlation, 2002-2003]{%
            \label{fig:pl01-03}
            \includegraphics[width=0.45\textwidth]{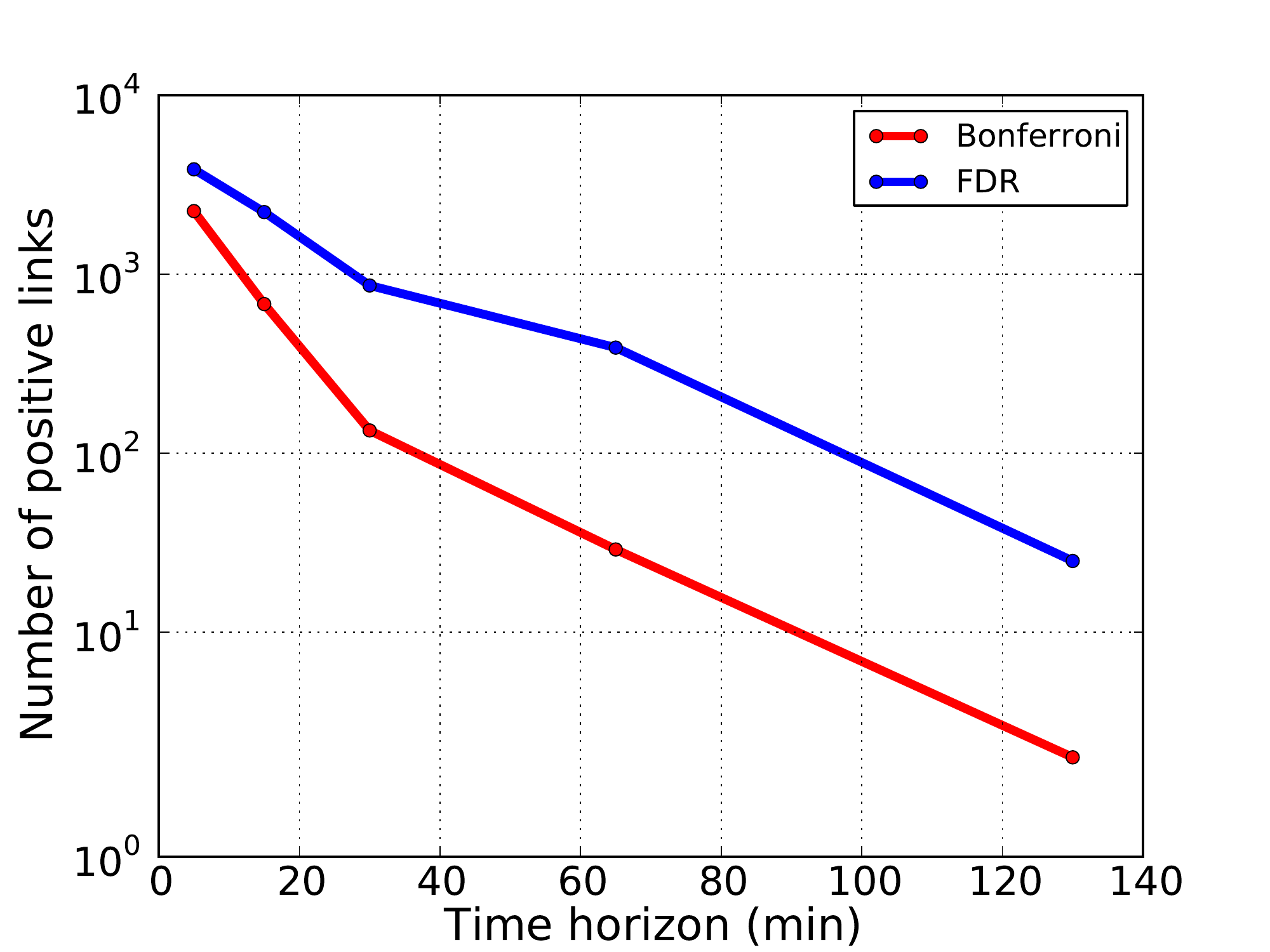}
            }
        \subfigure[Links of negative correlation, 2002-2003]{%
            \label{fig:nl01-03}
            \includegraphics[width=0.45\textwidth]{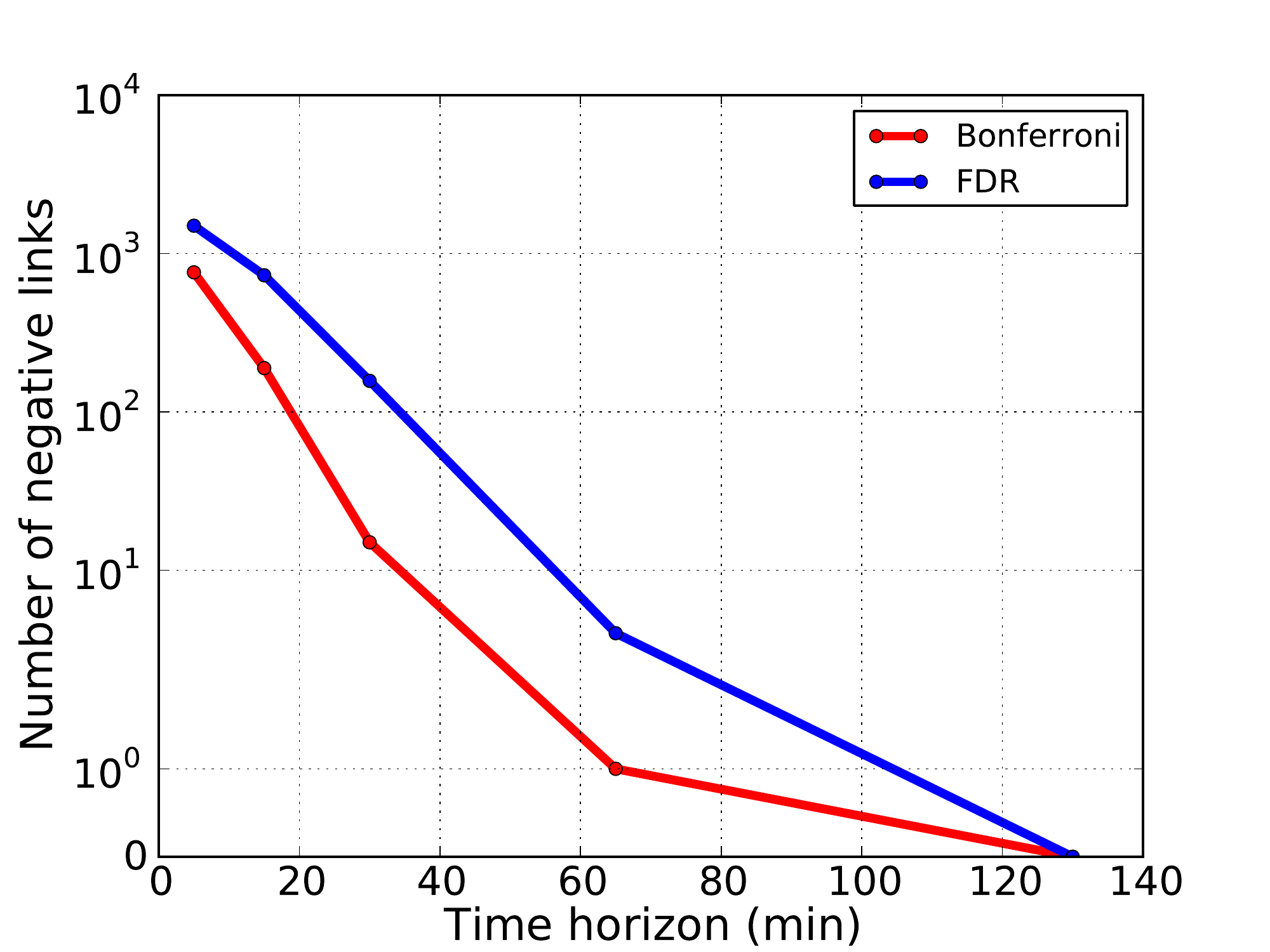}
            }}
            \centerline{
        \subfigure[Links of positive correlation, 2011-2012]{%
            \label{fig:pl11-12}
            \includegraphics[width=0.45\textwidth]{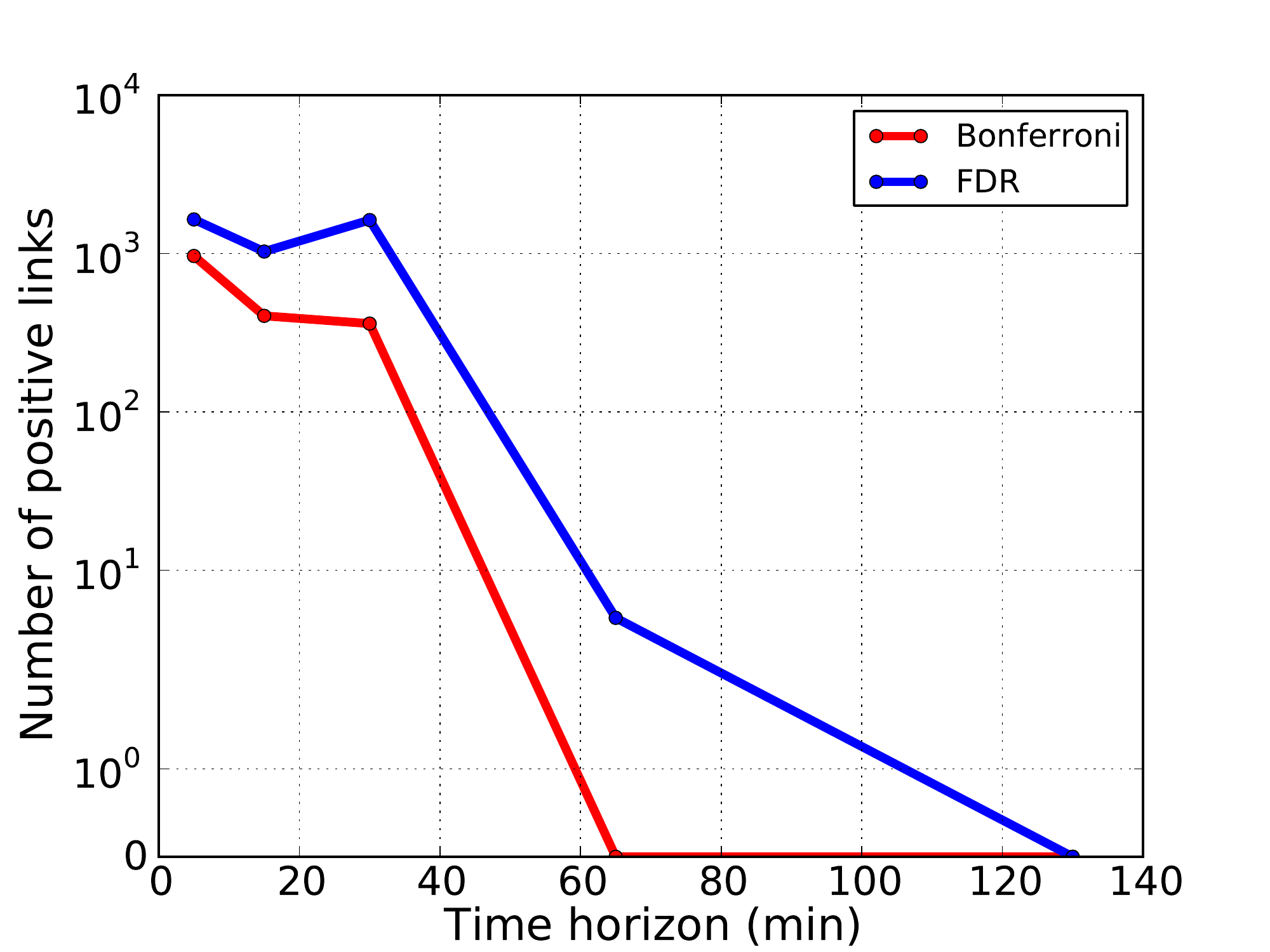}
            }
        \subfigure[Links of negative correlation, 2011-2012]{%
            \label{fig:nl11-12}
            \includegraphics[width=0.45\textwidth]{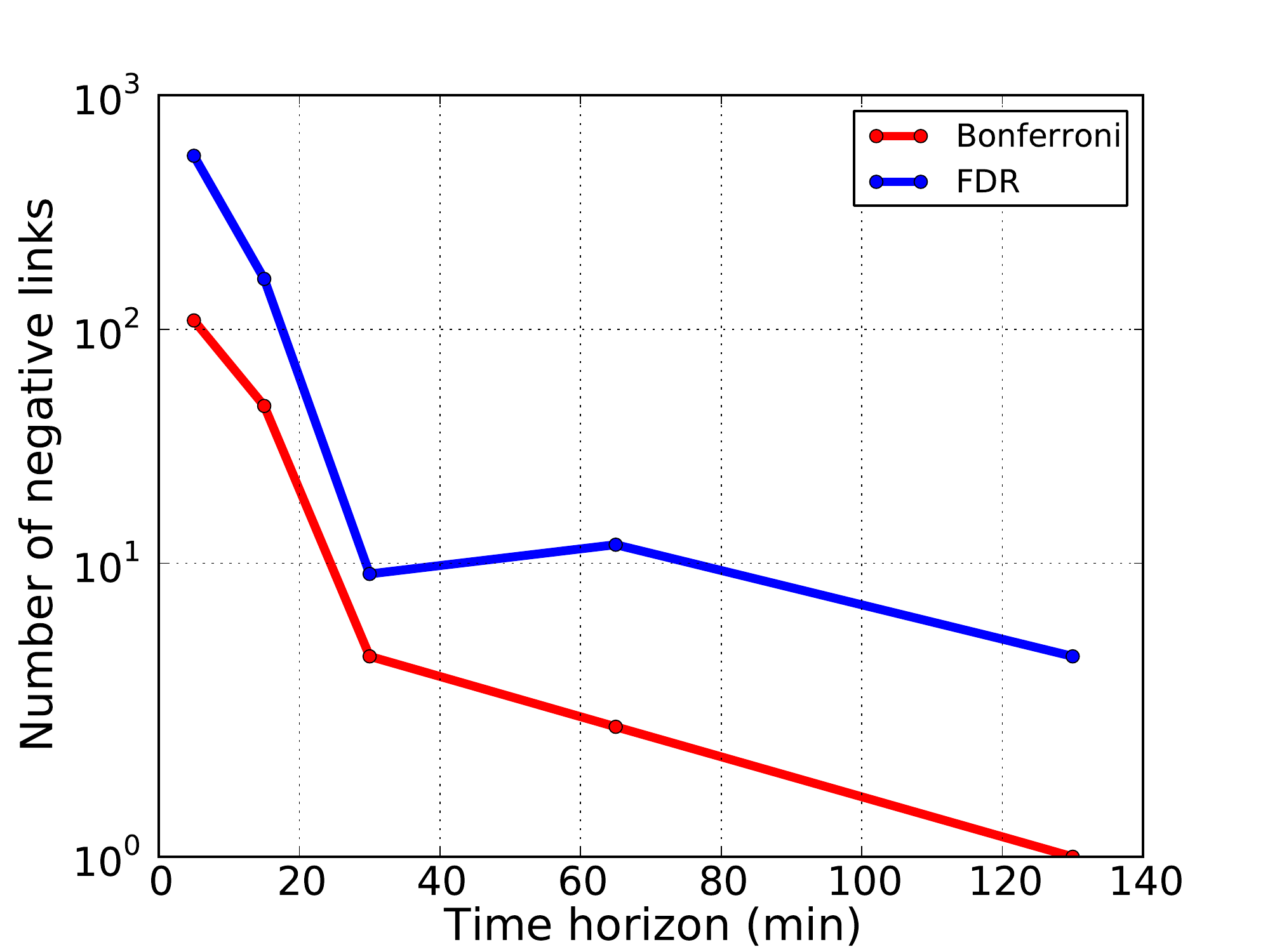}
            }}
    \end{center}
    \caption{%
        Plots of the number of positive and negative validated links for both Bonferroni and FDR lagged correlation networks. The decrease in number of validated links for increasing time horizon is apparent in both the 2002-2003 and 2011-2012 datasets. The vertical axis is presented on a logarithmic scale that is linearized near zero.
     }%
    \label{fig:number_links}
\end{figure}

It is clear visually that the validated links of positive correlation vastly outnumber
the validated links of negative correlation. We plot the number of validated links in
both the Bonferroni and FDR networks for the 2002--2003 and 2011--2012
datasets in Fig.~\ref{fig:number_links}, where the decrease in number of all
validated links for increasing time horizon is apparent. Note that for a
given time horizon we usually validate more links in the 2002--2003
dataset than in the 2011--2012 dataset. This suggests that there has
been an increase in market efficiency over the past decade. We revisit this idea in subsequent portions of this paper, where we study the properties of the network in- and out-degree distributions and the characterization of three-node motifs.

We also explore how the number of validated links decreases for a fixed
time horizon $h$ but a changing time lag. We build a lag $l$ into the
lagged correlation matrix (\ref{eqn:corr_matrix}) by excluding the last
$l$ returns of each trading day from matrix $A$ and the first $l$
returns of each trading day from matrix $B$. Thus the present analysis
uses $l=1$. In Appendix C we plot the decreasing number of validated
links with increasing $l$ for $h=15$ minutes.

\begin{figure}[h]
    \begin{center}
	 \centerline{
        \subfigure[2002-2003]{%
            \label{fig:lc01-03}
            \includegraphics[width=0.5\textwidth]{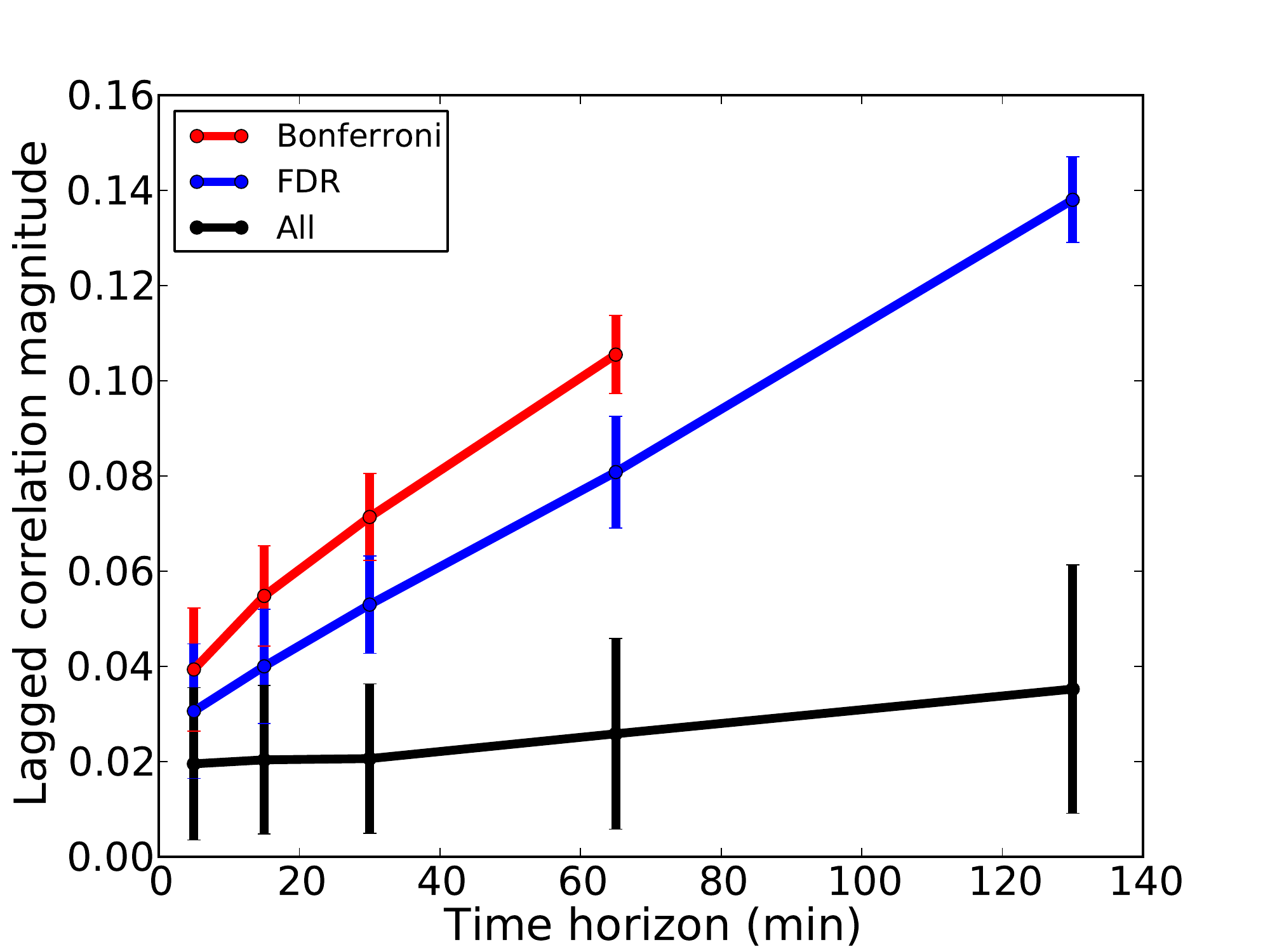}
            }
        \subfigure[2011-2012]{%
            \label{fig:lc11-12}
            \includegraphics[width=0.5\textwidth]{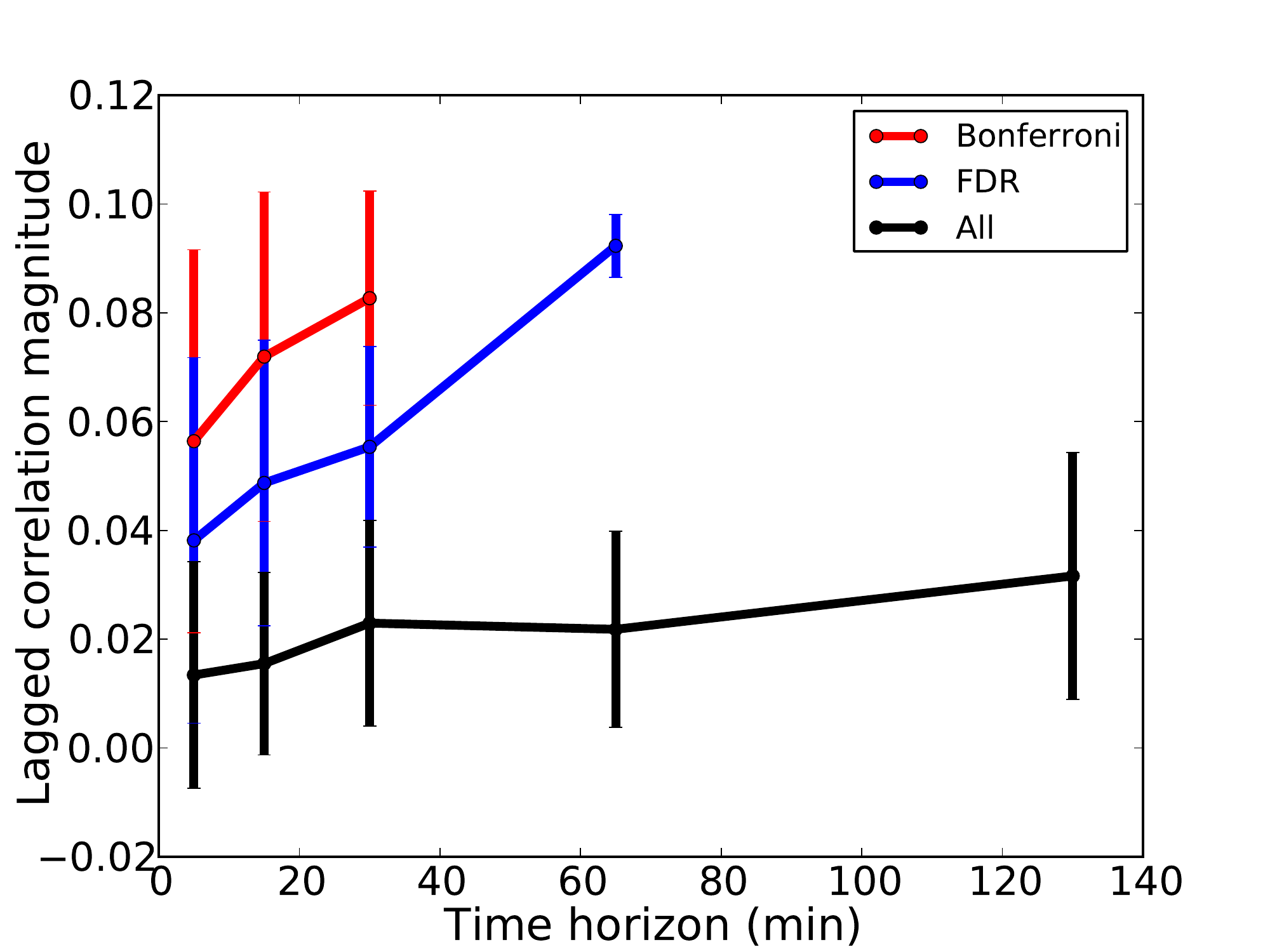}
            }}
    \end{center}
    \caption{%
        Average magnitude (absolute value) of lagged correlation coefficients filtered in Bonferroni and FDR networks. Magnitudes appear to grow with increasing time horizon of return sampling. Error bars represent plus-or-minus one standard deviation. Results are displayed only for networks containing at least five links.
     }%
    \label{fig:lagged_coeffs}
\end{figure}

We must also measure the extent to which the number of validated
lead-lag relationships can be disentangled from the strength of those
relationships.  Figure~\ref{fig:lagged_coeffs} thus shows plots of the
average magnitude of lagged correlation coefficients selected by the
Bonferroni and FDR networks. Although we validate more links at small
time horizons, we note that the average magnitude of the selected
coefficients tends to decrease. At short time horizons $h$ we correlate
time series of comparatively large length $T$, narrowing the distribution of entries in the shuffled lagged correlation matrix $\widetilde{C}$ and gaining statistical
power. We are thus able to reject the null hypothesis even for lead-lag
relationships with a modest correlation coefficient.

\begin{figure}[h]
    \begin{center}
	 \centerline{
        \subfigure[In-degree distributions of FDR networks]{%
            \label{fig:og}
            \includegraphics[width=0.5\textwidth]{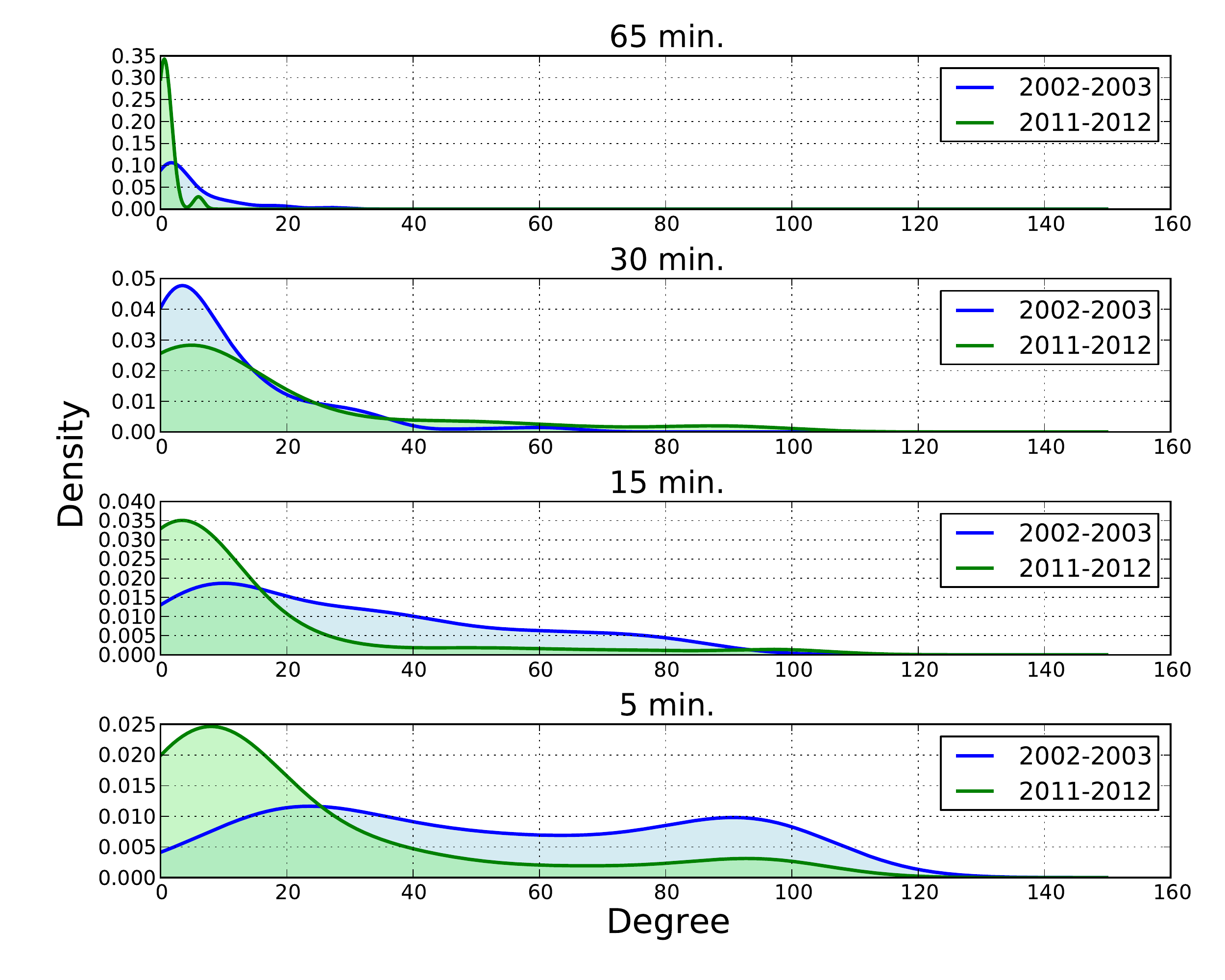}
            }
        \subfigure[Out-degree distributions of FDR networks]{%
            \label{fig:ig}
            \includegraphics[width=0.5\textwidth]{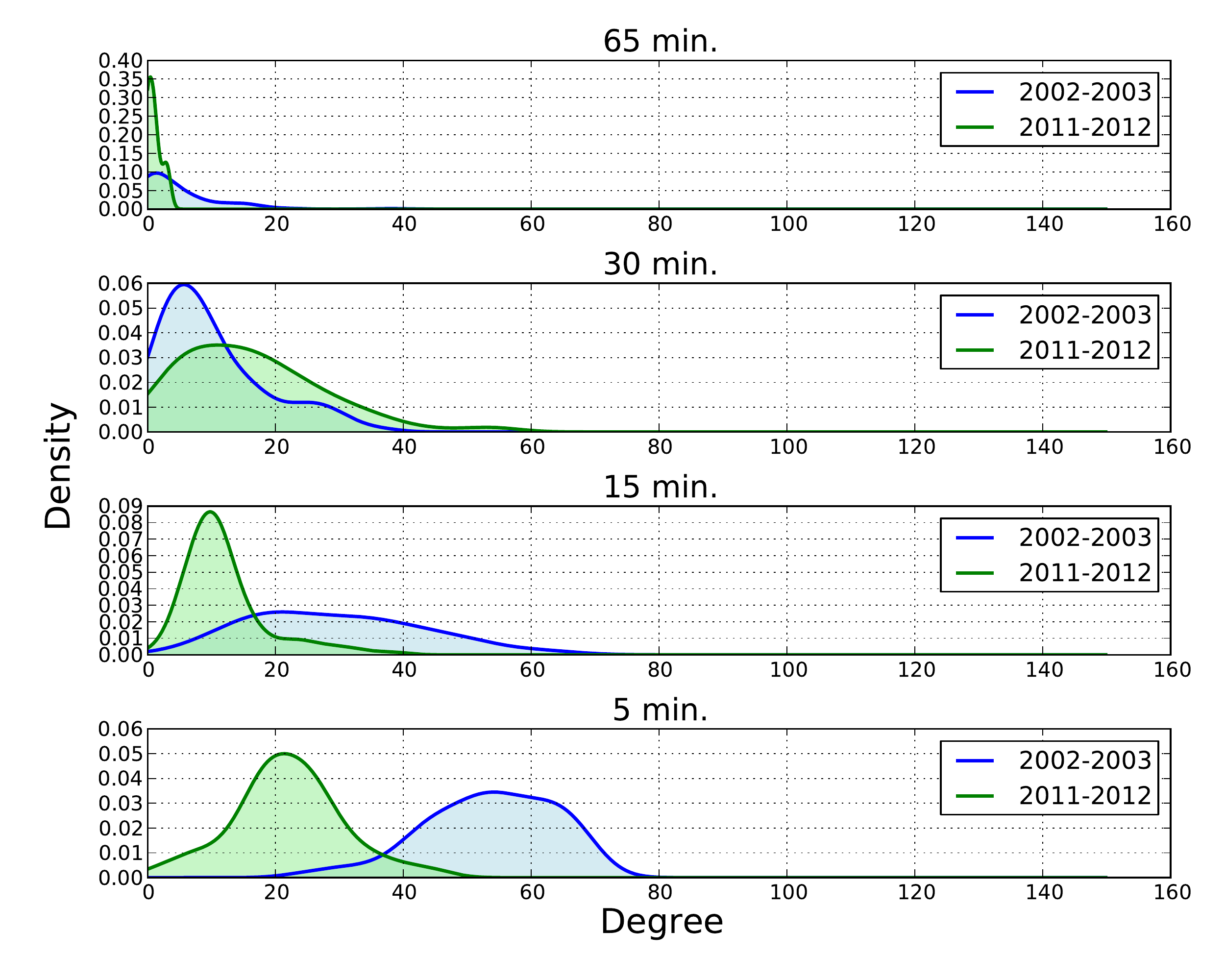}
            }}
    \end{center}
    \caption{%
        In- and out-degree distributions for FDR networks from 2002--2003 (blue) and 2011--2012 (green). Smoothed distributions are obtained using a kernel density estimate with a Gaussian kernel. With the exception of the $h=30$ minute in-degree distributions, at each of the 30 min., 15 min., and 5 min. time horizons the distributions from 2002--2003 and 2011--2012 are statistically distinct ($p<0.05$, all test statistics $W>130$, two-tailed two-sample Wilcoxon rank-sum tests, Bonferroni correction applied).
     }%
    \label{fig:degrees}
\end{figure}

Finally, in Fig.~\ref{fig:degrees} we characterize the topologies
of the statistically-validated networks by studying the properties of
their in-degree and out-degree distributions. We make two
observations. First, we note that both the in-degree and out-degree
distributions appear more homogeneous in the 2002--2003 period than the
2011--2012 period, i.e., the 2011--2012 data exhibit large
heterogeneities, particularly in the in-degree distributions, in which
many nodes have small degrees but few nodes have very large degrees, as
can be seen in the extended tails of the distributions. Second, we
observe that in both the 2002--2003 and 2011--2012 data there are more
nodes with large in-degrees than out-degrees. Although few individual
stocks have a strong influence on the larger financial market, it
appears that the larger financial market has a strong influence on many
individual stocks, especially at short time horizons.

\begin{figure}[h]                                      
\begin{center}
\centerline{
 \includegraphics[scale=.6]{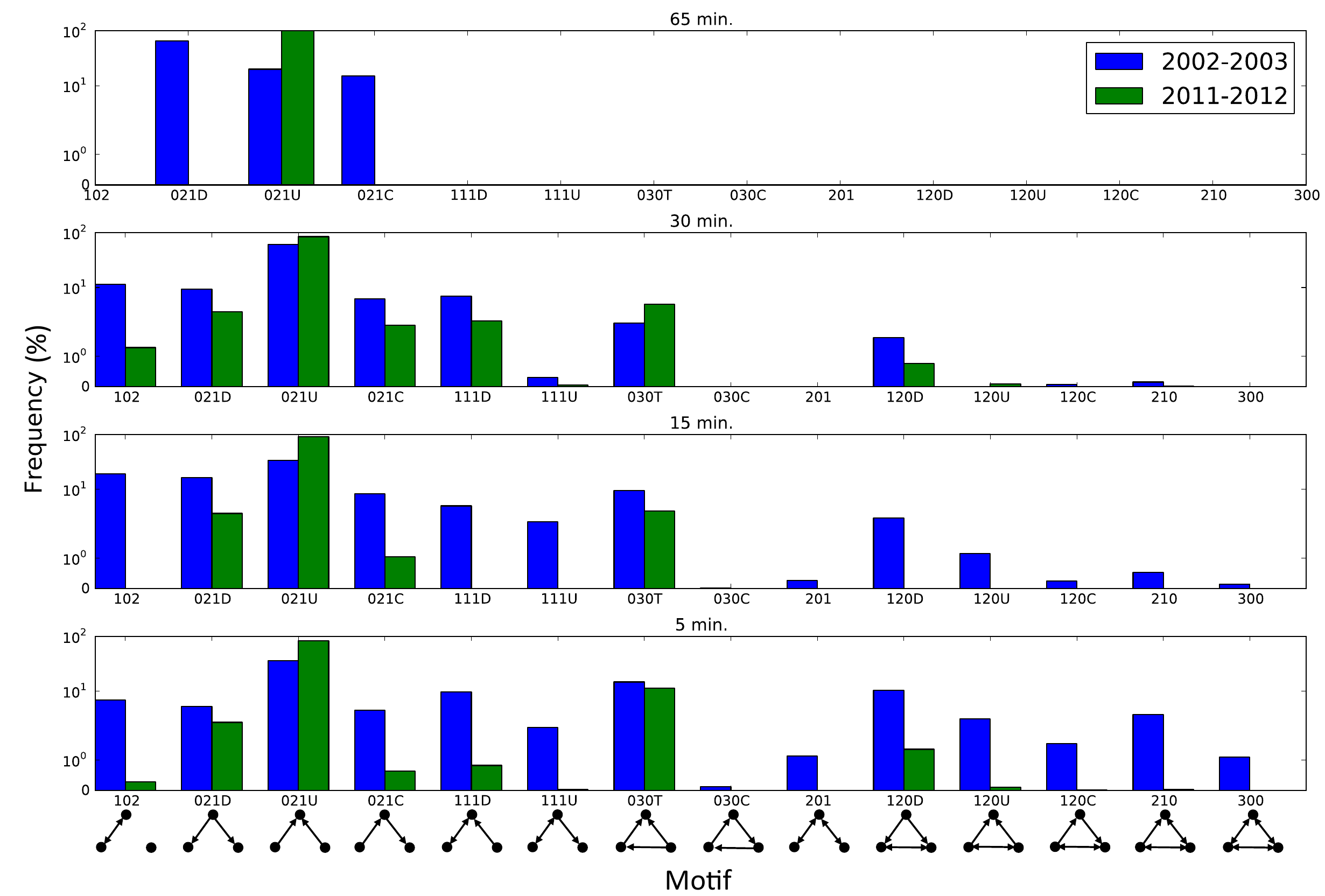}}
 \end{center}
\caption{Percentage occurrence of all 14 possible directed three-node motifs with more than one link in Bonferroni networks. The total number of such motifs in 2002-2003 are 40 ($h=65$ min.), 1,296 ($h=30$ min.), 17,545 ($h=15$ min.), and 92,673 ($h=5$ min.). In 2011-2012 these counts are 1 ($h=65$ min.), 9,171 ($h=30$ min.), 13,303 ($h=15$ min.), and 35,405 ($h=5$ min.).}
  \label{fig:motifs}
\end{figure}

We further investigate this point by studying the relative occurrence of
three-node network motifs in the Bonferroni networks
\citep{milo2002network}. We find that, of all motifs featuring more than
one link, the ``021U'' motif (two nodes influencing a common third node)
occurs frequently in the recent data, and in fact occurs in over 80\% of
node triplets having more than one link between them for time
horizons greater than $h=130$ minutes. In the 2002--2003 data this
motif is also the most common at every time horizon except $h =65$
minutes. Figure~\ref{fig:motifs} plots the occurrence frequencies of
these motifs. These features can be related to the information
efficiency of the market. In the 2011--2012 dataset we find a dominant
motif in which a large number of stocks influence only a few other
stocks. Predictive information regarding a given stock, therefore, tends
to be encoded in the price movements of many other stocks and so is
difficult to extract and exploit. In contrast, the distributions of
degrees and motifs in the 2002--2003 data are more homogeneous. Although
there are more nodes with large in-degrees, there are also more nodes
with large out-degrees. If a stock has a large out-degree, its price
movements influence the price movements of many other stocks. These
sources of exploitable information have all but disappeared over the
past decade.

\section{Synchronous correlation networks}

To construct synchronous correlation networks using the methodology
described in Sec.~\ref{sec:methods}, we use the unfiltered columns of
$R$ as our time series such that each entry $C_{m,n}$ of the empirical
correlation matrix is the Pearson correlation between columns $m$ and
$n$ of $R$. We then independently shuffle the columns of $R$, without
replacement, when constructing the surrogated time series. We find that
with the same significance threshold of $p=0.01$, in 2011-2012 both the Bonferroni
and FDR networks are almost fully connected, with well over 4500 of the
$N(N-1)/2=4950$ possible links validated in all networks over all time
horizons. Our method is thus quite sensitive to the presence or absence
of correlations between time series.

\begin{figure}[h]
    \begin{center}
	 \centerline{
        \subfigure[Epps curves for 2002-2003 and 2011-2012 data.]{%
            \label{fig:epps}
            \includegraphics[width=0.5\textwidth]{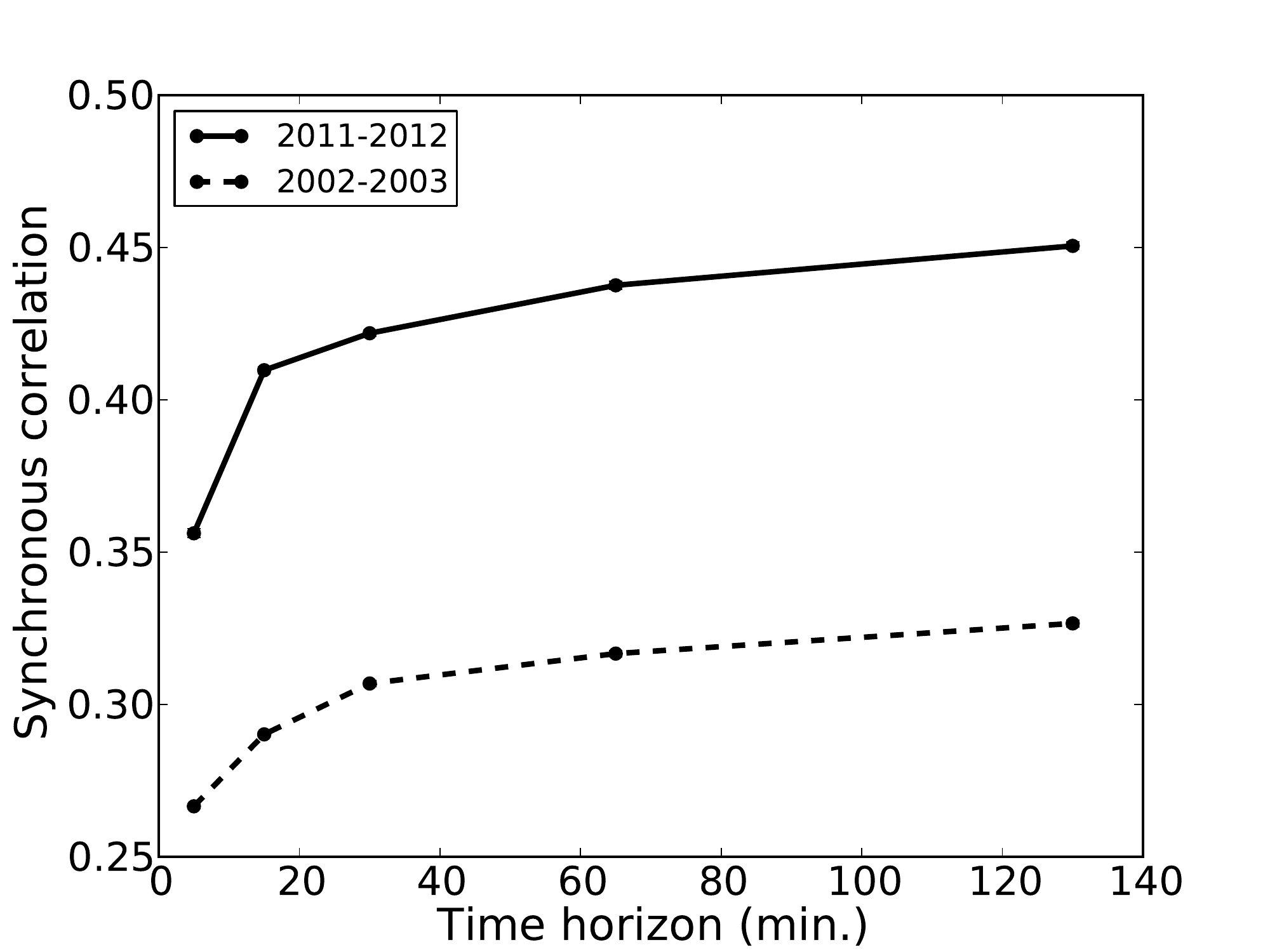}
            }
        \subfigure[Distributions of correlations at a 15 minute time horizon.]{%
            \label{fig:sync_corr_hist}
            \includegraphics[width=0.5\textwidth]{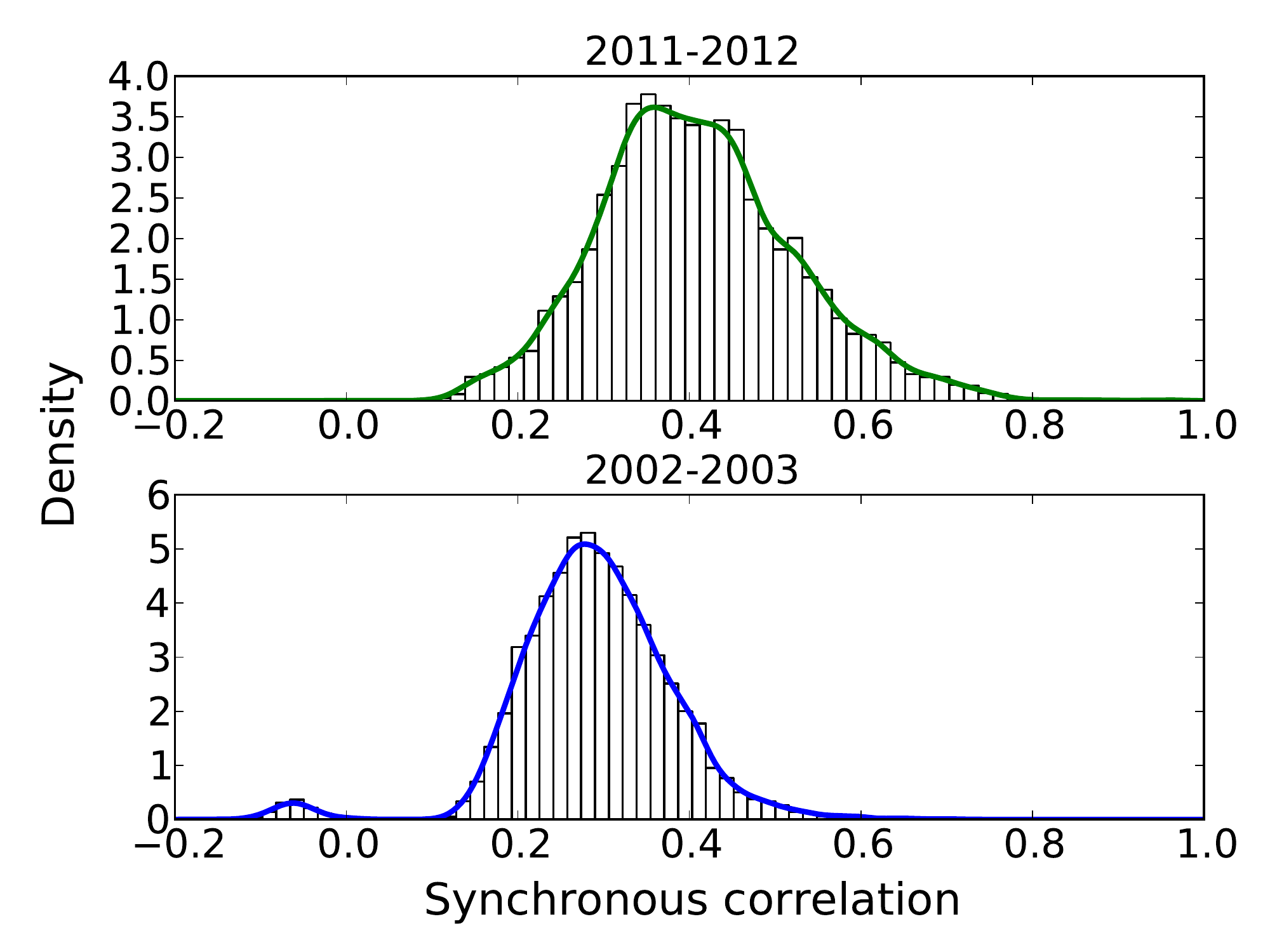}
            }}
    \end{center}
    \caption{ (a) Plot of mean synchronous correlation coefficients in
      both 2002-2003 and 2011-2012 data. Error bars represent
      plus-or-minus one standard deviation of the mean. (b) Histograms
      of correlation coefficients for returns sampled at a 15 minute
      time horizon. Solid curves show kernel density estimates using a
      Gaussian kernel. Depicted distributions are statistically distinct ($p<0.001$, test statistic $W=19601856$, two-tailed two-sample Wilcoxon rank-sum test).}
    \label{fig:sync_coeffs}
\end{figure}

Figure~\ref{fig:epps} plots the empirical synchronous correlations
against time horizon for all stocks considered in both datasets. We see
a clear increase in the magnitude of these coefficients as the time
horizon grows, a phenomenon known as the Epps Effect
\citep{epps1979comovements,tumminello2007correlation}. It is known that lagged correlations may in part contribute to this effect \citep{toth2009epps}. The extent of this contribution is an active area of investigation \citep{tumminello2014}. The synchronous
correlations are also significantly higher in the recent data,
suggesting that, despite the increased efficiencies shown in
Fig.~\ref{fig:number_links}, there is also an increase in co-movements
in financial markets since 2003, heightening the risk of financial
contagion (see for example \citep{song2011evolution,Kenett2012a}).

Figure \ref{fig:sync_corr_hist} shows the distribution of correlation
coefficients at $h=15$ minutes for both 2002--2003 and 2011--2012
datasets. We observe a slightly bi-modal distribution of synchronous
correlation coefficients in the 2002--2003 data across all time horizons
$h$. Most coefficients are positive, but there is also a small number of
negative coefficients among these high market capitalization
stocks. This quality disappears in the 2011--2012 data, and all
correlation coefficients are positive.

\section{Discussion}

In this paper, we propose a method for the construction of
statistically validated correlation networks. The method is applicable
to the construction of both lagged (directed) and synchronous
(undirected) networks, and imposes no topological
constraints on the networks. The sensitivity of the method to small deviations from the null hypothesis of uncorrelated returns makes it less
useful for studying the synchronous correlations of stocks, as these equities tend to display a considerable degree
of correlation and we validate almost all possible links in the
network. The method is apt, however, for the study of lagged correlation
networks. We are able to adjust the sensitivity of the method with our
choice of $p$-value and protocol for multiple comparisons. Here we show
that, with the conservative Bonferroni correction and $p$-value=0.01, we
are able to compare changes in network connectivity with increasing
return sampling frequency between old and new datasets. The primary
drawback to our method is its computational burden, which grows as
$\mathcal{O}(N^4)$ for $N$ time series.

We find that for timescales longer than one hour, significant lead-lag
relationships that capture return and information spill-over virtually
disappear. For timescales smaller than 30 minutes, however, we are able to validate
hundreds of relationships. According to the efficient market hypothesis
there can be no arbitrage opportunities in informationally-efficient
financial markets. However, lagged correlations may not be easily
exploitable due to the presence of market frictions, including
transaction costs, the costs of information processing, and borrowing
constraints.

Between the time periods 2002--2003 and 2011--2012, the synchronous
correlations among these high market capitalization stocks grow
considerably, but the number of validated lagged-correlation
relationships diminish. We relate these two behaviors to an increase in
the risks of financial contagion and an increase in the informational
efficiency of the market, respectively. We find that networks from both
periods exhibit asymmetries between their in-degree and out-degree
distributions. In both there are more nodes with large in-degrees than
large out-degrees, but in the 2011--2012 data, nodes with large
in-degrees are represented by the extended tails of the degree
distribution and, in contrast, the 2002--2003 distribution exhibits a
greater uniformity. A comparison between in-degree and out-degree distributions shows that nodes with high in-degree are much more likely than nodes with high out-degree, especially for the 2011--2012 data. This evidence is also interpreted in terms of informational efficiency of the market. Indeed a large out-degree of a stock implies that knowledge of its return, at a given time, may provide information about the future return of a large number of other stocks. On the other hand, a large in-degree of a stock indicates that information about its return at a given time can be accessed through the knowledge of past returns of many stocks. There are also many more nodes with large out-degrees in the 2002--2003 data than in the 2011--2012 data. We relate these observations to an increased
information efficiency in the market. Such an interpretation is also supported by the analysis of three-node motifs, which shows an apparent dominance of motif 021U with respect to all the others.

In the future, we could extend this work by incorporating a prediction
model to measure the degree to which the information contained in these
validated networks is exploitable in the presence of market
frictions. We could also investigate the characteristics of nodes
belonging to different industries, as well as the presence of intraday seasonalities \citep{allezbouchaud2011,tumminello2014}. Such features are potentially
relevant to prediction models. Finally, although our analysis restricts
itself to using the Pearson product-moment correlation, other measures,
such as a lagged Hayashi-Yoshida estimator
\citep{hayashi2005covariance}, could be used to probe correlations at
the smallest (inter-trade) timescales while minimizing the problem of
asynchronous trades.

\medskip
 
\section*{Acknowledgments.~~}

We thank Viktoria Dalko for useful conversations and insights, and her help with the data. CC, DYK, and HES wish to thank ONR (Grant N00014-09-1-0380, Grant N00014-12-1-0548),
DTRA (Grant HDTRA-1-10-1- 0014, Grant HDTRA-1-09-1-0035), and NSF (Grant CMMI 1125290). M.T. and R.N.M. acknowledge support from the INET research project NetHet ``New Tools in Credit Network Modeling with Heterogenous Agents''. R. N. M. acknowledge support from the FP7 research project CRISIS ``Complexity Research Initiative for Systemic InstabilitieS''.

\medskip

\section*{Author contributions.~~}

All authors contributed equally to this manuscript.

\vspace{-1mm}

\vspace{-1mm}

\bibliographystyle{rQUF}

\section*{Additional Information.~~}
The authors declare no competing financial interests.

\appendices
\section{Probability of a false positive link ~~}
The one-tailed $p$-value associated with positive correlations
represents the probability of observing a correlation between two
elements, $i$ and $j$, that is larger than or equal to the one observed,
$\rho_{obs}$, under the null hypothesis that $i$ and $j$ are
uncorrelated,
\begin{equation}
\text{$p$-value}(\rho_{obs})=P(\rho>\rho_{obs}).
\end{equation}
Our objective in the paper is to select all the correlations with a
$p$-value smaller than a given univariate statistical threshold $q_0$,
e.g., $q_0=0.01$ or $q_0=0.05$, corrected for multiple hypothesis
testing through the Bonferroni correction, that is, divided by the total
number of tests, $N^2$ in our case (where $N=100$ is the number of
stocks). The question is: \emph{what is the probability that a
  correlation with a p-value $p$ larger or equal to $p_0=q_0/N^2$ is
  (wrongly) indicated as a statistically significant one according to
  the shuffling method?}. Operatively, \emph{what is the probability
  that, over the $Q=k\,N^2$ independent replicates of the data, a
  correlation between $i$ and $j$ larger than the observed one has never
  been observed?}

If we set the $p$-value, $p$, of $\rho_{obs}$ equal to $\frac{q}{N^2}$
(where $q$ is a quantity that ranges between $0$ and $N^2$) the question
is: what is the probability that, over $Q=k\,N^2$ independent draws
($Q=100\cdot N^2=10^6$ bootstrap replicates with our method) a value of
correlation larger than $\rho_{obs}$ is never obtained? This probability
is
\begin{equation}
P(\text{null}|p)=(1-p)^Q,
\end{equation}
where ``null" indicates the event that a value of correlation larger
than $\rho_{obs}$ has never been obtained over $Q=k\,N^2$ random
replicates of data. This probability can be used to calculate the
probability that $p=q/N^2$ is larger than or equal to $p_0=q_0/N^2$,
conditioned to the event that a value of correlation larger than
$\rho_{obs}$ has never been obtained over $Q=k\,N^2$ draws. This is done
using Bayes' rule, under the assumption that the marginal distribution
of $p$-value $p$ is uniform in $[0,1]$, i.e., the density function is
$f(p)=1$ and then, integrating over $p$,
\begin{equation}
P(p \ge p_0|\text{null})=\int_{p_0}^{1}\frac{P(\text{null}|p) f(p)}{P(\text{null})} dp=\int_{p_0}^{1}(Q+1) (1-p)^Q dp = (1-p_0)^{Q+1},
\end{equation}
where we used the fact that $P(\text{null})=\int_{0}^1 P(\text{null}|p) f(p) dp$ $=\frac{1}{Q+1}$. 
In our method, $k=100$, and the sample size is $N=100$. Therefore 
\begin{equation}
P(p \ge p_0|\text{null}) =\left(1-\frac{q_0}{N^2}\right)^{k\,N^2+1}\cong \left(1-\frac{q_0}{N^2}\right)^{k\,N^2} \cong e^{-k\,q_0}.
\end{equation}
It is interesting to note that, as soon as the level of statistical
significance is corrected through the Bonferroni correction ($p_0 =
q_0/N^2$), where $q_0$ is the univariate level of statistical
significance, and the number, $Q$, of independent replicates is a
multiple of the number of tests, $Q=k\,N^2$, the probability $P(p \ge
p_0|\text{null})$ is approximately independent of the sample size ($N$).

With our approximated method to estimate correlation p-values, the
probability that we select a positive correlation as a statistically
significant one at the confidence level
$p_0=q_0/N^2=0.01/100^2=10^{-6}$, while it is actually not significant
at that level of statistical confidence, is $P(q \ge
0.01|\text{null})=\frac{1}{e} \cong 0.368$. However, the probability
that a significant correlation according to our method has a $p$-value
larger then $0.05/N^2=0.05/100^2=5\cdot10^{-6}$ is already quite small:
$P(q \ge 0.05|\text{null})=\frac{1}{e^5} \cong 0.0067$. In other words,
if we obtain a validated network with 1,000 links, i.e., 1,000 validated
positive correlations according to our approximated method, we expect
that, on average, only 7 correlations will have a one-tailed $p$-value
larger than $0.05/100^2=5 \cdot 10^{-6}$.
 
\section{Comparison of the bootstrap method and an
  analytical one to calculate correlation p-values~~}

Here we compare (for a sub-set of our data) the number of significant
correlations obtained according to the presented bootstrap approach and
the number of significant correlations that we may have obtained relying
upon the analytical distribution of sample pair correlations of normally
distributed data.

If $x$ and $y$ are uncorrelated variables that follow a normal
distribution, then the probability density function of the sample
correlation coefficient, $r$, between $x$ and $y$ is
\citep{kenney1962mathematics}
\begin{equation}
f(r,T)=\frac{(1 - r^2)^{\frac{T-1}{2} - 2}}{B(\frac{1}{2},\frac{T-1}{2}-1)},
\end{equation}
where $T$ is the length of the sample and $B(q,p)$ is the Euler beta
function of parameters $q$ and $p$. Given a level of statistical
significance, $q_0/N^2$ (already corrected for multiple hypothesis
testing), $f(r,T)$ can be used to set a threshold for the correlation
value $\rho_{t}$ such that the probability $P(\rho >
\rho_{t})=\frac{q_0}{N^2}$ is
\begin{equation}
P(\rho > \rho_{t})=\int_{\rho_{t}}^{1}f(r,T) dr=\frac{q_0}{N^2}.
\end{equation}
According to this analysis, for a data sample of $N$ time series, each
one of length $T$, we can say that an observed correlation,
$\rho_{obs}$, is statistically significant if $\rho_{obs}>\rho_t$, where
$\rho_t$ is obtained by (numerically) solving the previous non linear
equation.

Table B1 shows the 2002--2003 dataset and reports the length of data
series used to calculate lagged correlations (column 1) at a given time
horizon (column 2), the quantity $\rho_t$ such that $P(\rho > \rho_{t})
= 0.01/N^2=10^{-6}$ (column 3), the number of validated positive
correlations (column 4), and the number of validated negative
correlations (column 5).

\begin{table}
\centering
\tbl{Threshold-correlation values and validated links according \\
to a normal distribution of returns}
{\begin{tabular}{@{}ccccc}
\toprule
$T$ & $h$ & $\rho_t$ & \#  pos. valid. & \# neg. valid \\
\hline
38,577 & 5 min & 0.0242 & 2,398 & 793 \\
12,525 & 15 min & 0.0425 &  754 & 212\\
6,012 & 30 min & 0.0613  & 158 & 19\\
2,505 & 65 min & 0.0948  & 43 & 3 \\
1002 & 130 min & 0.1496 & 3 & 0\\
\botrule
\end{tabular}}
\end{table}

Table B2 shows the number of validated positive correlations (i) according to
the shuffling method (column 3), (ii) according to the analytical method
discussed above (column 4), and (iii) common to both methods (column 5). The
results reported in the table show that the bootstrap method we used is
more conservative than the analytical method based on the assumption
that return time series follow a normal distribution. Indeed the number
of validated positive correlations according to the bootstrap method is
always smaller than the one obtained using the theoretical
approach. Furthermore, most of the correlations validated according to
the bootstrap method are also validated according to the theoretical
method.

\begin{table}
\centering
\tbl{Comparison between number of validated links according\\ to 1)
  bootstrap method and 2) a normal distribution of returns}
{\begin{tabular}{@{}ccccc}
\toprule
$T$ & $h$ & \#  pos. valid. (bootstrap) & \#  pos. valid. (normal dist.) & \# pos. valid (both) \\
\hline
38,577 & 5 min & 2,252 & 2,398 & 2,230 \\
12,525 & 15 min & 681 &  754 & 666\\
6,012 & 30 min &  134 & 158 & 131\\
2,505 & 65 min &  29 & 43 &  26\\
1002 & 130 min & 2  & 3 & 2\\
\botrule
\end{tabular}}
\end{table}
A similar discussion can be held about the validation of negative correlations.

\section{Effect of lag and time series length on validated links for a fixed time horizon}

\begin{figure}[h]
    \begin{center}
	 \centerline{
        \subfigure[Links of positive correlation]{%
            \label{fig:pos_lag}
            \includegraphics[width=0.5\textwidth]{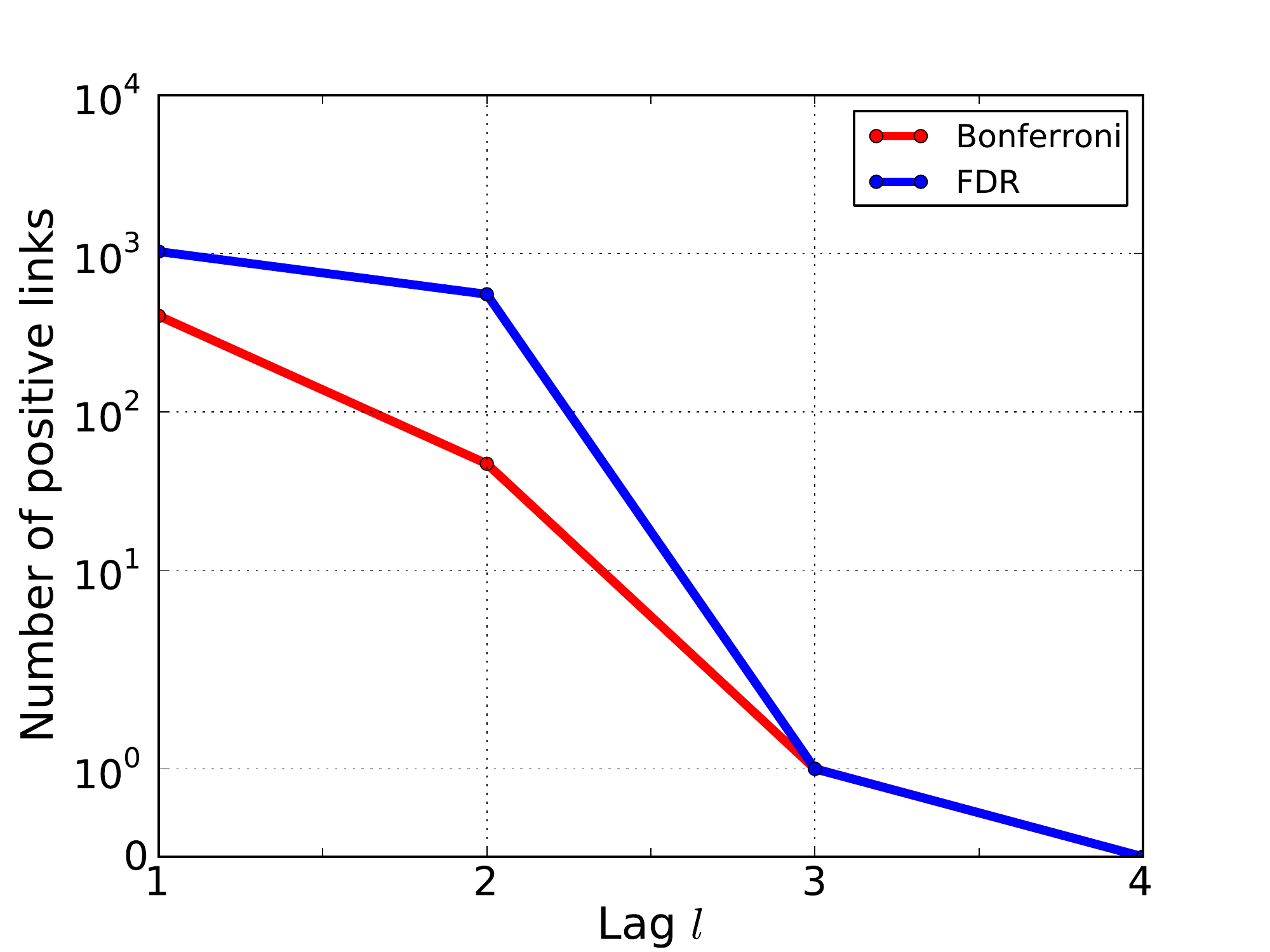}
            }
        \subfigure[Links of negative correlation]{%
            \label{fig:neg_lag}
            \includegraphics[width=0.5\textwidth]{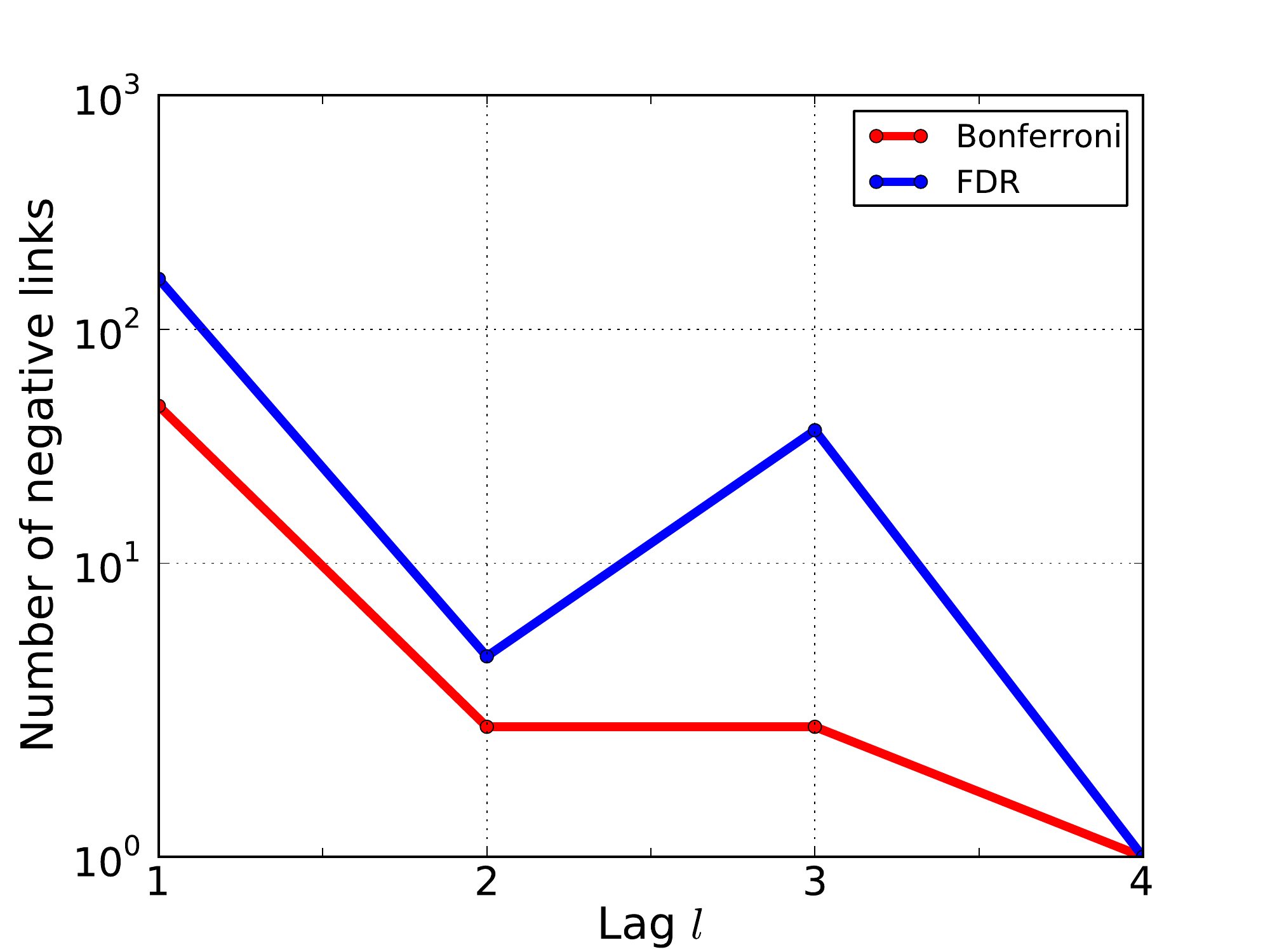}
            }}
    \end{center}
    \caption{ Numbers of positive and negative validated links for both
      Bonferroni and FDR correlation networks for varying lag
      $l$. Returns are sampled every $h = 15$ minutes from the 2011-2012
      data. }
    \label{fig:variable_lag}
\end{figure}

We explore how the number of validated links decreases when the time
horizon $h$ is fixed and the time lag variable $l$ increases. A lag $l$ is built into
the lagged correlation matrix (\ref{eqn:corr_matrix}) by excluding the
last $l$ returns of each trading day from matrix $A$ and the first $l$
returns of each trading day from matrix $B$. Thus the results presented
in the main text are restricted to $l =
1$. Figure~\ref{fig:variable_lag} plots the number of positive links and
negative links validated in the 2011--2012 data for $h=15$ minutes as
$l$ increases. Although for this $h$ the length $T$ of the time series
in $A$ and $B$ decrease by only $\approx 4$\% for each additional lag
$l$ (as each 390 minute trading day includes $390/15 - l = 26-l$
returns), we observe a sharp decrease in the number of validated links
as $l$ increases. The number of validated negative links is an order of
magnitude smaller than the number of positive links, so the small peak
in negative links at $l=3$ for the FDR network is likely an artifact of
noise.

\begin{figure}[h]
    \begin{center}
	 \centerline{
        \subfigure[Links of positive correlation]{%
            \label{fig:pos_months}
            \includegraphics[width=0.5\textwidth]{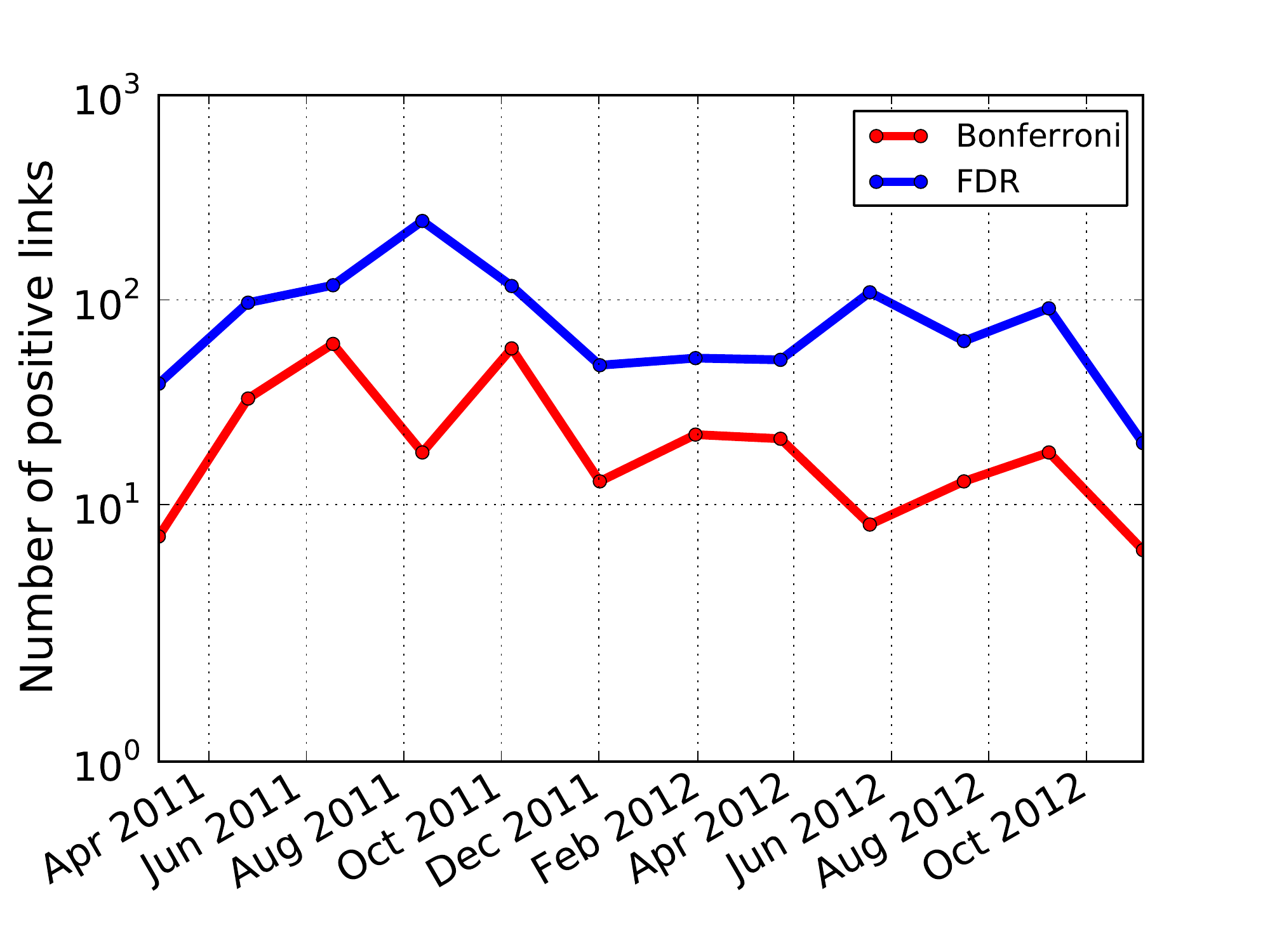}
            }
        \subfigure[Links of negative correlation]{%
            \label{fig:neg_months}
            \includegraphics[width=0.5\textwidth]{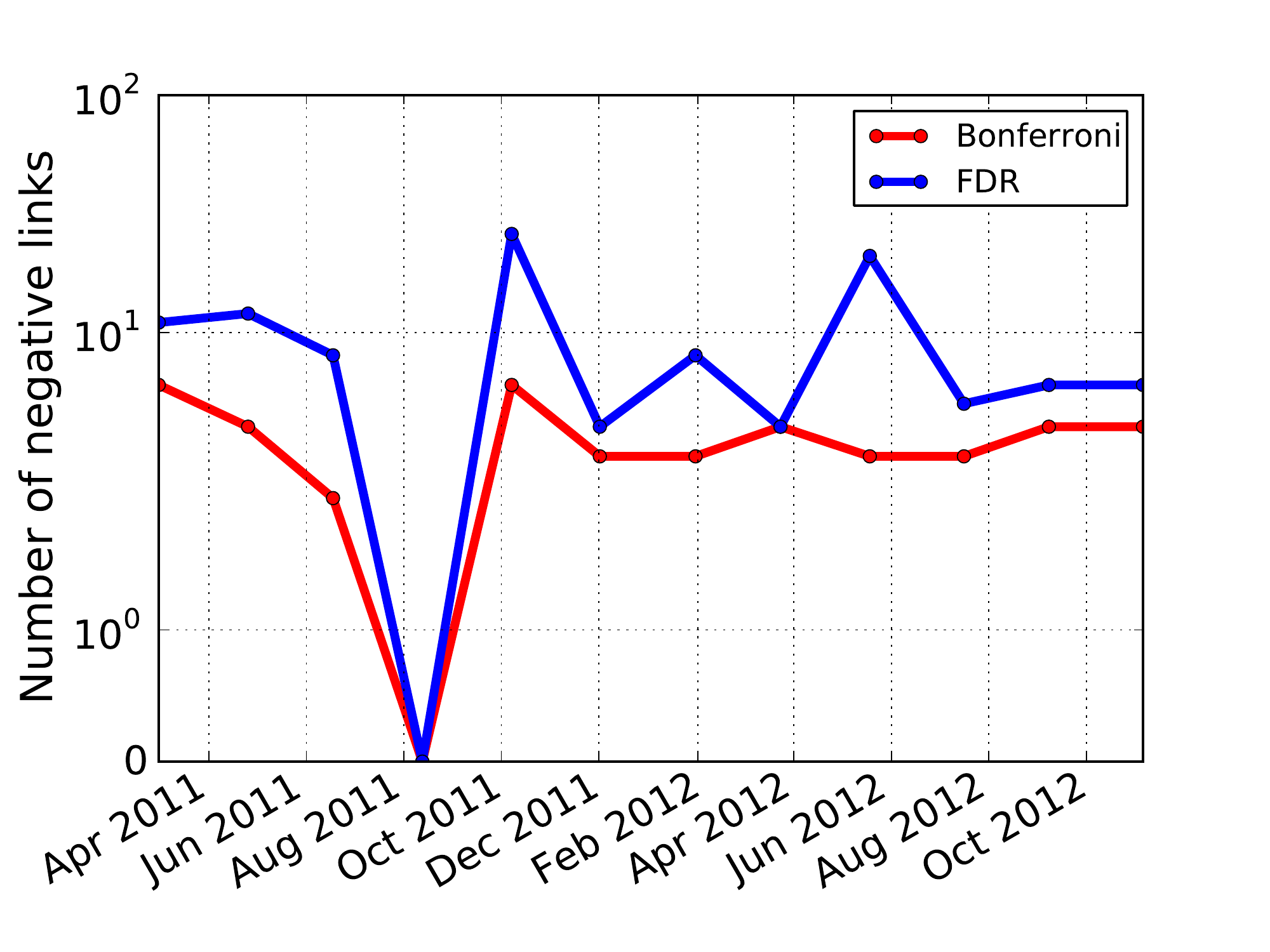}
            }}
    \end{center}
    \caption{%
        Numbers of positive and negative validated links for both Bonferroni and FDR lagged correlation networks for time series segments of length $T = 1004$ at $h = 15$ minutes. Horizontal axis gives date of the final return in each network.
     }%
    \label{fig:fixed_T}
\end{figure}

We also investigate the effect of the time series length $T$ on the numbers of validated links. For $h=15$ minutes, we partition the entire 2011-2012 time series into segments of length $T=1004$, as this is the length of the time series for the longest time horizon considered ($h=130$ minutes). For each segment we generate the lagged correlation network using $10^6$ surrogate time series, as before. We find that the union of all such Bonferroni networks consists of 124 distinct links, 104 of which are positive and 20 of which are negative. Although this number is 27\% of the number of links validated in the $h=15$ minute network that was not partitioned ($T=12,550$), it stands in contrast to the single link that was validated in the $h=130$ minute Bonferroni network using the entire time period. The number validated in each partition is shown in Figure \ref{fig:fixed_T}. We can thus safely conclude that decreasing the time horizon $h$ provides information independent of the increased time series length $T$.

\end{document}